\documentclass[12pt]{article}
\usepackage{amssymb}
\usepackage{amsfonts}
\usepackage{amsmath}
\usepackage[mathscr]{eucal}
\usepackage{amssymb}
\usepackage{amsthm}
\usepackage{bbold}
\usepackage{bm}
\usepackage{graphicx}
\usepackage{caption}
\usepackage{hyperref}
\usepackage{cite}

\theoremstyle{plain}

\numberwithin{equation}{section}
\newcommand{\be}{\begin{equation}}
\newcommand{\ee}{\end{equation}}
\newcommand{\bea}{\begin{eqnarray}}
\newcommand{\eea}{\end{eqnarray}}
\newcommand{\nn}{\nonumber \\}
\newcommand{\lb}{\label}

\begin{document}

\begin{titlepage}

\vspace*{0.2cm}

\renewcommand{\thefootnote}{\star}
\begin{center}

{\LARGE\bf  ${\cal N}{=}\,4$ supersymmetric multiparticle systems}\\

\vspace{0.5cm}

{\LARGE\bf  based on indecomposable multiplets}\\

\vspace{1.5cm}
\renewcommand{\thefootnote}{$\star$}

{\large\bf Sergey\,Fedoruk${\,}^{a)}$},\,\,\,{\large\bf Evgeny\,Ivanov${\,}^{a)\,b)}$},\,\,\,{\large\bf Stepan\,Sidorov${\,}^{a)}$}
 \vspace{0.8cm}

{\small {\it ${\,}^{a)}$ Bogoliubov Laboratory of Theoretical Physics, JINR, 141980 Dubna, Moscow Region, Russia} \\
\vspace{0.2cm}
{\it ${\,}^{b)}$ Moscow Institute of Physics and Technology, 141700 Dolgoprudny, Moscow Region, Russia} \\
\vspace{0.3cm}
{\tt fedoruk@theor.jinr.ru, \, eivanov@theor.jinr.ru, \, sidorovstepan88@gmail.com}}\\
\end{center}
\vspace{0.2cm} \vskip 0.6truecm \nopagebreak

   \begin{abstract}
\noindent We construct new multiparticle models of $\mathcal{N}=4$ supersymmetric mechanics with spin degrees of freedom by employing nonlinear indecomposable
supermultiplets ${\bf (1,4,3){\supset\hspace{-1.1em}+}(4,4,0)}$.
These systems are proper deformations of those associated with the standard irreducible $d=1$, $\mathcal{N}=4$ multiplets.
In this way we find a deformation of $\mathcal{N}=4$ supersymmetric U$(2)$-spin rational Calogero system invariant under
$d=1$ superconformal group OSp$(4|2)$. One more deformed model reproduces $\mathcal{N}=4$ supersymmetric U$(2)$-spin hyperbolic Calogero system,
up to a shift of the Hamiltonian by some U$(1)$ generators.
\end{abstract}
\vspace{1cm}

\begin{flushright}
{\it In Memory of Jerzy Lukierski (1936 - 2026)}
\end{flushright}

\vspace{1.5cm}

\phantom{Keywords: }

\newpage

\end{titlepage}

\setcounter{footnote}{0}

\setcounter{equation}0
\section{Introduction}

Growing interest in supersymmetric Calogero models is caused by their close relationships with black hole physics \cite{Claus98, GT1998} and AdS/CFT correspondence \cite{Maldacena}. Construction of the first supersymmetric
extension of the multiparticle Calogero mechanical system with $d=1$, ${\cal N}=2$ superconformal symmetry OSp$(2|2)$ \cite{FM1990}\footnote{Models of superconformal mechanics
were proposed in the pioneering works
\cite{AP1983,FR1984} for $\mathcal{N}=2$ case and in \cite{IKL1989} for $\mathcal{N}=4$.} initiated a lot of further studies on superconformal mechanics and the associated super Calogero systems (see, {\rm e.g.},
\cite{BHKV1993, Wyl2000, BGIK0212, GLP0708, LN1509,KLS2204} and \cite{FIL1112} for a more complete list of references). The Calogero-like models with internal spin degrees of freedom were pioneered in \cite{Poly1991}  as
gauged matrix models (see also a review \cite{Poly0607}). As was shown in \cite{FIL0812}, $\mathcal{N}=1,2,4$ supersymmetrizations of such models can be accomplished by applying the superfield gauging procedure
\cite{DI0605,DI0611}. An important feature of this approach is the use of the off-shell matrix $d=1$ superfields carrying linear irreducible supermultiplets of the underlying supersymmetries. Its one more crucial ingredient
is the simultaneous consideration of both dynamical and the so-called semi-dynamical superfields (accommodating spin degrees of freedom).

Supersymmetric $d=1$ models with dynamical and semi-dynamical (spin) degrees of freedom can be also built on the indecomposable (not fully reducible or long)
supermultiplets \cite{Toppan1006,Toppan1204,Hubsch1310,CasimirEnergy,IS1509}. The
characteristic feature of models with indecomposable supermultiplets is the presence of an additional parameter allowing to treat
such models as certain deformations of models with irreducible linear supermultiplets.

A recent study of nonlinear indecomposable (long) $\mathcal{N}=8$ supermultiplets \cite{IS2512} gave rise to the off-shell formulation of $\mathcal{N}=8$ supersymmetric model with spin degrees
of freedom described earlier in \cite{FI2402} in terms of $\mathcal{N}=4$ superfields.\footnote{This ${\cal N}=8$ model was shown in \cite{KhKN2408} to be invariant under OSp$(8|2)$ superconformal symmetry.
The further studies along these lines resulted in discovering new superconformal $d=1$ systems \cite{KN2411}.}
In ref. \cite{IS2512}, $\mathcal{N}=8$ superfield description of these new  indecomposable multiplets was given  at the ``kinematical'' off-shell
level (constraints and component contents),  while the problem of building  the relevant manifestly $\mathcal{N}=8$ invariant superfield Lagrangians remained unsolved.
Only the component and $\mathcal{N}=4$ superfield actions for these supermultiplets were partially worked out.

As a step towards constructing the complete  ${\cal N}=8$ superfield description of such systems, in the present  paper we consider $\mathcal{N}=4$ counterpart of the nonlinear
indecomposable $\mathcal{N}=8$ supermultiplet of ref. \cite{IS2512} and construct the related models of ${\cal N}=4$ mechanics.

As a characteristic example, it is instructive to recall first the simpler linear indecomposable (long) $\mathcal{N}=4$ multiplet introduced in \cite{SS2410}.

Off-shell multiplets of $d=1$, $\mathcal{N}=4$ supersymmetry can be conveniently presented in the harmonic superspace formulation of $\mathcal{N}=4$ supersymmetric mechanics \cite{IL0307,IS2112}.
As was shown in \cite{SS2410}, the simplest example of indecomposable $\mathcal{N}=4$ multiplet is a system of the scalar real superfield $\mathscr{X}$ and the analytic  superfield ${\cal W}^{++}$,
which are subject to the conditions \footnote{Here we use the notations $\mathscr{X}$ and ${\cal W}^{++}$ instead of $X_{\kappa}$ and $V^{++}$ of ref. \cite{SS2410}.
Some necessary information on $d=1, {\cal N}=4$ harmonic superspace is collected in Appendix \ref{AppA}.}
\be\label{143+341}
{\rm (a)}\;D^{+}_{\alpha}D^{+\alpha}\mathscr{X} = 4i\varkappa\,\mathcal{W}^{++},\qquad {\rm (b)}\;D^{++}\mathscr{X}=0\,,
\ee
and
\be\label{341}
D^{+}_{\alpha}\mathcal{W}^{++}=0\,,\qquad D^{++}\mathcal{W}^{++}=0\,,
\ee
where $\varkappa$ is a constant.
The off-shell constraints \eqref{341} are those defining the standard irreducible  multiplet ${\bf (3,4,1)}$ of $d=1, \,\mathcal{N}=4$ supersymmetry\cite{IL0307}.
The meaning of the constraints \eqref{143+341} is different for the cases $\varkappa =0$ and $\varkappa \neq 0$:
\begin{itemize}
    \item When $\varkappa=0$, the constraints\eqref{143+341} define the mirror multiplet ${\bf (1,4,3)}$ \cite{IS2112}.
    Therefore, the superfields  $\mathscr{X}$ and ${\cal W}^{++}$ describe two irreducible multiplets and we face the fully reducible case;
    \item When $\varkappa\neq 0$, the conditions \eqref{143+341} linearly relate $\mathscr{X}$ and $\mathcal{W}^{++}$, the analytic superfield $\mathcal{W}^{++}$ still representing the
    irreducible  multiplet ${\bf (3,4,1)}$. In this case, the superfield $\mathscr{X}$ describes a quotient of the original set $(\mathscr{X}, \mathcal{W}^{++})$ over an irreducible subset
    corresponding to the multiplet ${\bf (3,4,1)}$ and so we encounter an indecomposable multiplet in this case.
\end{itemize}
We denote this multiplet for an arbitrary $\varkappa$  as a semidirect sum ${\bf (1,4,3){\supset\hspace{-1.1em}+}(3,4,1)}$. Sending $\varkappa \to 0$ leads to a direct sum
of irreducible multiplets ${\bf (1,4,3)}$ and ${\bf (3,4,1)}$, {\it i.e.} the multiplet becomes fully reducible in this limit.

From another point of view \cite{IvSmi1991}, for $\varkappa \neq 0$
the unconstrained scalar superfield $\mathscr{X}$ with $8+8$ off-shell components can be considered as a prepotential
for the analytic superfield $\mathcal{W}^{++}$ describing the multiplet ${\bf (3,4,1)}$,
$\mathcal{W}^{++} \sim D^{+}_{\alpha}D^{+\alpha}\mathscr{X}$. Half of the original total $16$ components of $\mathscr{X}$ are pure gauge in such an interpretation.
Then, picking up a gauge-invariant action for $\mathscr{X}$ and finally passing to the appropriate Wess-Zumino gauge
leave us with $4+4$ off-shell components of $\mathcal{W}^{++}$.
This means that the gauge invariant actions  of this kind provide just another form of the off-shell description of the multiplet ${\bf (3,4,1)}$.
A non-trivial option is to start with gauge-non-invariant actions and this was the basic idea of ref. \cite{SS2410}.
In this case all components of the set $(\mathscr{X}, \mathcal{W}^{++})$  contribute to the action and it describes the indecomposable
multiplet ${\bf (1,4,3){\supset\hspace{-1.1em}+}(3,4,1)}$ with $8+8$ off-shell degrees of freedom (or two irreducible multiplets ${\bf (1,4,3)}$ and ${\bf (3,4,1)}$ in the limit $\varkappa=0$).

Here we introduce a new indecomposable $d=1, \,\mathcal{N}=4$ multiplet corresponding to a semidirect sum ${\bf (1,4,3){\supset\hspace{-1.1em}+}(4,4,0)}$.
It is described by the superfields $\mathscr{X}$ (real) and $\mathcal{Z}^{+}, \bar{\mathcal{Z}\,}^{+}$ (complex) satisfying the constraints
\bea
&&D^{+}_{\alpha}D^{+\alpha}\mathscr{X} = 2\varkappa\,{\mathcal{Z}}^{+}\bar{\mathcal{Z}\,}^{+},\qquad D^{++}\mathscr{X}=0\,,\label{(eq-143)in} \\ [6pt]
&&({\rm a})\;D^{+}_{\alpha}{\mathcal{Z}}^{+}=0\,,\qquad
({\rm b})\;D^{++}{\mathcal{Z}}^{+}=0\,, \quad ({\rm and \; c.c.}). \label{(440)in}
\eea
The relevant component off-shell set matches the field content of the matrix models considered in \cite{FIL0812,FIL1902,F2007} and providing
$\mathcal{N}=4$ supersymmetric generalizations of spin Calogero systems on shell.
In contrast to the multiplet ${\bf (1,4,3){\supset\hspace{-1.1em}+}(3,4,1)}$, the new indecomposable $\mathcal{N}=4$ multiplet is nonlinear, because
the superfields $\mathscr{X}$ and $\mathcal{Z}^{+}$ are related by the nonlinear condition \eqref{(eq-143)in}.
In the limit $\varkappa \to 0$ we end up with two independent $\mathcal{N}=4$ superfields representing the irreducible multiplets ${\bf (1,4,3)}$ and ${\bf (4,4,0)}$.
As distinct from the constraints \eqref{143+341}, eqs. \eqref{(eq-143)in} cannot be used anyway to express ${\mathcal{Z}}^{+}$ in terms of $\mathscr{X}$, so the latter cannot be treated as
a prepotential for ${\mathcal{Z}}^{+}$. The notable feature of the multiplet ${\bf (1,4,3){\supset\hspace{-1.1em}+}(4,4,0)}$,  in both the one-particle and the multiparticle cases,
is that the actions for its ${\bf (1,4,3)}$ part induce Wess-Zumino (WZ) type actions for the invariant subset ${\bf (4,4,0)}$, with the strength proportional  to the deformation parameter $\varkappa$.
No such a peculiarity takes place in the case of the multiplet ${\bf (1,4,3){\supset\hspace{-1.1em}+}(3,4,1)}$.

In the present paper, we consider both  the single-particle and multiparticle ${\cal N}=4$ models with spin degrees of freedom,  based on the multiplet
${\bf (1,4,3){\supset\hspace{-1.1em}+}(4,4,0)}$.
The plan of the paper is as follows. In Section 2 we discuss in more detail the constraints  \eqref{(eq-143)in}, \eqref{(440)in} as those defining a deformation of the irreducible multiplets
${\bf (1,4,3)}$ and ${\bf(4,4,0)}$ into the indecomposable one  ${\bf (1,4,3){\supset\hspace{-1.1em}+}(4,4,0)}$. We construct general superfield and component Lagrangians for
a single multiplet. In Section 3 we generalize the multiplet  ${\bf (1,4,3){\supset\hspace{-1.1em}+}(4,4,0)}$ to matrix superfields subject
to the local U$(n)$ gauge transformations. In Section 4 we construct a new gauged $\mathcal{N}=4$ superconformal multiparticle model with the bosonic sector
being a deformation of the rational Calogero system. In Section 5 we construct a simple multiparticle model providing a supersymmetrization of the hyperbolic Calogero-Sutherland system.
It proves to be equivalent to the previously known supersymmetrization, since the relevant Hamiltonians differ merely by a term proportional to some U$(1)$ generators commuting with the Hamiltonians.
Section 6 contains a summary of the results and indicates some possible areas of their implications.
In Appendix \ref{AppA}, we present necessary details of the harmonic  $d=1$, ${\cal N}=4$ formalism.

\section{Indecomposable multiplet (1,4,3)$\,{\supset\hspace{-0.95em}+}\,$(4,4,0)}

\subsection{Multiplet (4,4,0)}

The invariant subset of the indecomposable multiplet
${\bf (1,4,3){\supset\hspace{-1.1em}+}(4,4,0)}$ is the multiplet ${\bf (4, 4, 0)}$, which in the harmonic approach is described
by an analytic complex superfield ${\mathcal{Z}}^{+}(\zeta_{\mathrm{A}})$, $\zeta_{\mathrm{A}} := (t_{\mathrm{A}}, \theta^{+}_{\alpha}, u^\pm_i)$,
subject to the off-shell constraints \eqref{(440)in} \cite{IL0307}.
The constraints $({\rm a})$ and $({\rm b})$ in \eqref{(440)in} are, respectively, Grassmann and harmonic analyticity conditions
imposed with the help of the covariant derivatives defined in \eqref{D++D+}.
The conjugated superfield $\bar{\mathcal{Z}}^{+}(t_A, \theta^{+}_{\alpha}, u^\pm_i)$ obeys the same conditions:
\be\label{440c}
D^{+}_{\alpha}\bar{\mathcal{Z}}^{+}=0\,,\qquad
D^{++}\bar{\mathcal{Z}}^{+}=0\,.
\ee
The solution of eqs. \eqref{(440)in} and \eqref{440c} for the analytic superfields $\mathcal{Z}^{+}$ and $\bar{\mathcal{Z}}^{+}$ are given by the  following component expansions
\footnote{Note that $(\overline{u_i^\pm})= {}\mp u^{\mp i}$ for complex conjugation, $(u_i^\pm)^*={}\pm u_i^\mp$ for antipodal conjugation,
and $(\widetilde{\,u_i^\pm\,})=u^{\pm i}, (\widetilde{\,u^{\pm i}})=- u^{\pm}_ {i}$ for the ``tilde'' ($\sim\,$) conjugation. Respectively,
$(\overline{\theta_\alpha^i})=(\widetilde{\theta^i_{\alpha}}) = \theta^\alpha_i$\,, $\theta_\alpha^\pm=\theta_\alpha^i u_i^\pm$, $(\widetilde{\theta_\alpha^\pm})={}-\theta^{\pm\alpha}$.}
\be\label{ZbarZ}
{\mathcal{Z}}^{+}=z^{+} - \theta^{+}_{\alpha}\xi^{\alpha}+i\,\theta^{+}_{\alpha}\theta^{+\alpha}\dot{z}^{-},\qquad
\bar{\mathcal{Z}}^{+}=\bar{z}^{+}-\theta^{+\alpha}\bar{\xi}_{\alpha}+i\,\theta^{+}_{\alpha}\theta^{+\alpha}\dot{\bar z}^{-}
,\qquad \widetilde{\mathcal{Z}^{+}}=\bar{\mathcal{Z}}^{+},
\ee
where $z^{\pm} = z^{i}u^{\pm}_i$, $\bar{z}^{\pm}=\bar{z}_{i}u^{\pm i}$ and dot denotes the derivative with respect to the time variable. In expansions \eqref{ZbarZ}, the component fields
are bosonic ones
$z^{i}$, $\overline{\left(z^{i}\right)}=\bar{z}_{i}$ and fermionic ones
$\xi^{\alpha}$, $\overline{\left(\xi^{\alpha}\right)}=\bar{\xi}_{\alpha}$.
The supersymmetry transformations are realized on these component fields as:
\be\label{tr_440}
\delta z^{i}=\epsilon^{i}_{\alpha}\xi^{\alpha}\,,\quad \delta \bar{z}_{i}=-\,\epsilon_{i}^{\alpha}\bar{\xi}_{\alpha}\,,\qquad \delta \xi^{\alpha}=-\,2i\,\epsilon_{i}^{\alpha}\dot{z}^{i}\,,\quad\delta \bar{\xi}_{\alpha}=2i\,\epsilon^{i}_{\alpha}\dot{\bar{z}}_{i}\,,
\ee
where $\epsilon^{i}_{\alpha}={}\overline{\left(\epsilon_{i}^{\alpha}\right)}$ are $d=1, \mathcal{N}=4$ supertranslation parameters.

The role of the component fields in the specific supersymmetric $d=1$ models is determined by the type of the relevant Lagrangian.
In models with the dynamical multiplet ${\bf (4, 4, 0)}$, the component Lagrangian is of the second order in time derivative for fields $z^{i}$
and of the first order for fields $\xi^{\alpha}$. In the case of semi-dynamical multiplet ${\bf (4, 4, 0)}$, fields $\xi^{\alpha}$ are auxiliary ({\it i.e.}, have no kinetic terms),
while the Lagrangian for the fields $z^{i}$ is
of the first order in time derivative. In the present paper we will deal just  with the second option.

\subsection{Unifying the  multiplets (1,4,3) and (4,4,0)}

In this paper we deal with the so-called ``mirror'' multiplet ${\bf (1, 4, 3)}$ which is described in the harmonic approach by a real superfield $\mathscr{X}$ subject to the conditions $D^{+}_{\alpha}D^{+\alpha}\mathscr{X}
=0$ and $D^{++}\mathscr{X} =0$ \cite{DI0611,IS2112}.

Passing to the indecomposable multiplet that involves the multiplet ${\bf (1, 4, 3)}$ as the proper quotient over the multiplet ${\bf (4, 4, 0)}$
amounts to considering the superfield ${\mathscr{X}}$ subject to the deformed constraints \eqref{(eq-143)in}:
\be
D^{+}_{\alpha}D^{+\alpha}{\mathscr{X}} = 2\varkappa\,{\mathcal{Z}}^{+}\bar{\mathcal{Z}}^{+},\qquad D^{++} {\mathscr{X}} = 0\,,\nonumber
\ee
where the superfields ${\mathcal{Z}}^{+}$ and $\bar{{\mathcal{Z}}}^{+}$
satisfy the off-shell conditions \eqref{(440)in} and \eqref{440c}.

The common solution of eqs. \eqref{(eq-143)in} and \eqref{(440)in}, \eqref{440c} reads:
\bea
\mathscr{X}&\!=\!&x-\theta^{-}_{\alpha}\psi^{i\alpha}u^{+}_i+\theta^{+}_{\alpha}\psi^{i\alpha}u^{-}_i+\theta^{-}_{(\alpha}\theta^{+}_{\beta)}N^{\alpha\beta}
+i\,\theta^{-}_{\alpha}\theta^{+\alpha}\dot{x}-i\,
\theta^{+}_{\alpha}\theta^{+\alpha}\theta^{-}_{\beta}\dot{\psi}^{i\beta}u^{-}_i
\nonumber \\ [5pt]
&\!\!&-\, \frac{\varkappa}{2}\left(\theta^{+}_{\alpha}\theta^{+\alpha}z^{-}\bar{z}^{-}-
\theta^{+}_{\alpha}\theta^{-\alpha}z^{+}\bar{z}^{-}-
\theta^{+}_{\alpha}\theta^{-\alpha}z^{-}\bar{z}^{+}+\theta^{-}_{\alpha}
\theta^{-\alpha}z^{+}\bar{z}^{+}\right)
\nonumber \\ [5pt]
&\!\!&-\,\frac{\varkappa}{2}\Big[\theta^{-}_{\alpha}\theta^{-\alpha} \theta^{+}_{\beta}\left(z^{+}\bar{\xi}^{\beta}-\bar{z}^{+}\xi^{\beta}\right)+\theta^{+}_{\alpha}\theta^{+\alpha} \theta^{-}_{\beta}\left(z^{-}\bar{\xi}^{\beta}-\bar{z}^{-}\xi^{\beta}\right)\Big]
\nonumber \\ [5pt]
&\!\!&-\, \frac{\varkappa}{4}\,\theta^{+}_{\alpha}
\theta^{+\alpha}\theta^{-}_{\beta}\theta^{-\beta}
\Big[2i \big(\dot{z}^{-}\bar{z}^{+}+z^{+}\dot{\bar{z}}^{-} \big)
-\xi^{\gamma}\bar{\xi}_{\gamma}\Big]\,,
\lb{X-sol}
\eea
where the component fields
$x=\overline{\left(x\right)}$ and $N_{\alpha\beta}={}-\overline{\left(N^{\alpha\beta}\right)}$ are bosonic, whereas $\psi_{i\alpha}=\overline{\left(\psi^{i\alpha}\right)}$ is fermionic.
The fields $x$, $\psi^{i\alpha}$ and $N^{\alpha\beta}$ are transformed under $d=1, {\cal N}=4$ supersymmetry as:
\be\label{tr-143-l}
\begin{array}{c}
\delta x = \epsilon_{i\alpha}\psi^{i\alpha},\qquad
\delta \psi^{i\alpha} = \epsilon^i_{\beta}N^{\alpha\beta}+i\,\epsilon^{i\alpha}\dot{x}+\varkappa\,\epsilon^{\alpha}_{j}z^{(i}\bar{z}^{j)},
\\ [5pt]
\delta N^{\alpha\beta} = 2i\,\epsilon^{i(\alpha}\dot{\psi}^{\beta)}_i+\varkappa\,\epsilon^{i(\alpha}\left(z_i\bar{\xi}^{\beta)}+\bar{z}_i\xi^{\beta)}\right).
\end{array}
\ee
In the limit $\varkappa = 0$ one recovers the ``mirror'' version of the multiplet ${\bf (1, 4, 3)}$.

It is useful to summarize what we have for the moment. In the system described by the superfields $\mathscr{X}$ and $\mathcal{Z}^+, \bar{\mathcal{Z}}^+$, the component fields of $\mathcal{Z}^+, \bar{\mathcal{Z}}^+$
(multiplet ${\bf (4, 4, 0)}$) transform through themselves under supersymmetry transformations,
whereas the component expansion of the superfield  $\mathscr{X}$ (multiplet ${\bf (1, 4, 3)}$) involves the fields of the latter multiplet together with
those of $\mathcal{Z}^+$. The same concerns the supersymmetry transformations of the components of $\mathscr{X}$. This is just the reason why we use
the notation ${\bf (1,4,3){\supset\hspace{-1.1em}+}(4,4,0)}$ for such a deformed ${\cal N}=4$  multiplet.
It should be pointed out that the constraints \eqref{(eq-143)in} and the supersymmetry transformations \eqref{tr-143-l} are nonlinear.
So we encounter the nonlinear indecomposable $d=1$ supermultiplet ${\bf (1,4,3){\supset\hspace{-1.1em}+}(4,4,0)}$
which was not considered before.

The related important point is that the constraints \eqref{(eq-143)in} lead to the appearance of the component fields of superfield $\mathcal{Z}^+$ in the $\theta$-expansion of $\mathscr{X}$.
As a result, some simple (e.g., looking ``free'')  Lagrangians of $\mathscr{X}$
could yield in fact  a non-trivial interacting Lagrangians in which, among the component fields,
there will be present the component fields of the superfield $\mathcal{Z}^+$.
These properties of the indecomposable multiplets is a useful tool of constructing new non-trivial ${\cal N}=4$  theories with interacting component fields.
In order to avoid an unnecessary broadening of the text, we will not discuss this type of Lagrangians in detail here, and move directly on models with gauge symmetry.
The simplest non-gauge models can be directly obtained from the gauge-invariant ones in the limit of vanishing gauge fields.

One more important feature of the set of constraints \eqref{(eq-143)in} and \eqref{(440)in}, \eqref{440c} is that they respect
invariance not only under $d=1$, ${\cal N}=4$ super Poincar\'e transformations, but also under superconformal symmetry ${\rm D}(2,1;\alpha)$.

The conformal supersymmetry coordinate transformations
\bea
&&  \delta^{\prime} t_{\rm A}
= 2i\,\eta^{-\alpha}\theta^{+}_{\alpha}\,t_{\rm A}\,,\qquad
\delta^{\prime} u^{+}_{i}=2i\alpha\,\eta^{+\alpha}\theta^{+}_{\alpha}u^{-}_{i}\,, \qquad
\delta^{\prime} u^{-}_{i}=0 \,,
\nn  [5pt]
&&
\delta^{\prime} \theta^{+}_{\alpha}= \eta^{+}_{\alpha}\,t_{\rm A}+i\left(1+\alpha\right)\eta^{-}_{\alpha}\theta^{+}_{\beta}\theta^{+\beta},
\nn [5pt]
&&\delta^{\prime} \theta^{-}_{\alpha}= \eta^{-}_{\alpha}\,t_{\rm A}-i\left(1+\alpha\right)
\eta^{+}_{\alpha}\theta^{-}_{\beta}\theta^{-\beta}+2i\left(1+\alpha\right)
\eta^{-}_{\alpha}\theta^{+}_{\beta}\theta^{-\beta}
-2i\alpha\,\eta^{-\beta}\theta^{+}_{\beta}\theta^{-}_{\alpha}\,,\lb{tr-d21a}
\eea
where $\eta^{\pm}_{\alpha} = \eta^{i}_{\alpha}u^{\pm}_{i}$, lead
to the following transformations of the covariant derivatives:
\be
\begin{array}{rcl}
\delta^{\prime} D^{++} &=& -\,2i\alpha\,\eta^{+\alpha}\theta^{+}_{\alpha}D^{0}\,,
\\ [5pt]
\delta^{\prime} D^{+\alpha} &=& 2i\alpha\,\eta^{-\beta}\theta^{+}_{\beta}D^{+\alpha}-2i\left(1+\alpha\right)
\left(\eta^{+}_{\beta}\theta^{-\alpha}-\eta^{-}_{\beta}\theta^{+\alpha}\right)
D^{+\beta}\,.
\end{array}
\ee
If the involved superfields are transformed as
\be\label{conf}
\delta^{\prime}{\mathscr{X}} = 2i\left(1+\alpha\right)\left(\eta^{-\alpha}\theta^{+}_{\alpha}-\eta^{+\alpha}\theta^{-}_{\alpha}\right){\mathscr{X}}\,,\qquad
\delta^{\prime}{\mathcal{Z}}^{+}=2i\alpha\,\eta^{-\alpha}\theta^{+}_{\alpha}\,{\mathcal{Z}}^{+}\,,
\ee
eqs.  \eqref{(eq-143)in}, \eqref{(440)in} and \eqref{440c} are covariant under the whole $d=1, {\cal N}=4$ superconformal symmetry ${\rm D}(2,1;\alpha)$.

This property of indecomposable supermultiplets allows one to use them for constructing new superconformal $d=1,{\cal N}=4$ systems.

\subsection{Gauging one-particle system}

The superfield gauging procedure was elaborated in \cite{DI0605, DI0611}. For ${\cal N}=4$ case it essentially uses the formalism of
$d=1$, ${\cal N}=4$ harmonic superspace \cite{IL0307} which is $d=1$ reduction of $4D$, ${\cal N}=2$ harmonic superspace \cite{HSS1984}.
The basic object is the gauge harmonic superfield ${\cal V}^{++}$ defined on the analytic subspace of $d=1, {\cal N}=4$ harmonic superspace.
It respects twice as less supersymmetries compared to its $4D$, ${\cal N}=2$ precursor. Due to this difference, the Wess-Zumino gauge
of ${\cal V}^{++}$ carries no degrees of freedom except for a ``topological'' one \cite{DI0605}. This remaining component
was exploited as a non-propagating matrix ``gauge field'' in \cite{FIL0812}.

Basic equations \eqref{(440)in}, \eqref{440c} and \eqref{(eq-143)in}
are covariant with respect to the rigid ${\rm U}(1)$ transformations:
\be\lb{u1-tr}
\mathscr{X}^{\prime}=\mathscr{X}\,,\qquad {\mathcal{Z}}^{+}{}^{\prime}=e^{i\lambda}\,{\mathcal{Z}}^{+}\,,\qquad
\bar{{\mathcal{Z}}}^{+}{}^{\prime}=e^{-i\lambda}\,\bar{{\mathcal{Z}}}^{+}\,,
\ee
where $\lambda$ is a real parameter. We start from the system with symmetry \eqref{u1-tr}
and promote $\lambda$ to an analytic superfield. That is,
$D^{+}_{\alpha}\lambda=0$ and $\lambda=\lambda(\zeta_{\mathrm{A}})$ is a function of the coordinates
$\zeta_{\mathrm{A}}=(t_{\mathrm{A}}, \theta^{+}_{\alpha}, u^\pm_i)$ of the analytic harmonic superspace.
With respect to this local symmetry, the derivative $D^{+}_{\alpha}$ is covariant on its own  and
does not require any fermionic gauge connection. To covariantize the harmonic covariant derivative $D^{++}$,
we introduce the analytic superfield $\mathcal{V}^{++}(\zeta_{\mathrm{A}})$, with the following gauge transformation law
\be\lb{u1-tr-V}
{\mathcal{V}}^{++}{}^{\prime}={\mathcal{V}}^{++}-D^{++}\lambda\,.
\ee
Then the covariant counterpart of the constraints \eqref{(eq-143)in} and \eqref{(440)in}, \eqref{440c} is provided by the set
\be\label{eq-143-r}
D^{+}_{\alpha}D^{+\alpha}\mathscr{X} = 2\varkappa\,{\mathcal{Z}}^{+}\bar{{\mathcal{Z}}}^{+}\,,\qquad D^{++} \mathscr{X} = 0\,,
\ee
\be\label{GaugeInvConstraints}
\begin{array}{l}
D^{+}_{\alpha}{\mathcal{Z}}^{+}=0\,,\qquad
\big(D^{++}+i\,{\mathcal{V}}^{++}\big){\mathcal{Z}}^{+}=0\,,
\\ [7pt]
D^{+}_{\alpha}\bar{{\mathcal{Z}}}^{+}=0\,,\qquad
\big(D^{++}-i\,{\mathcal{V}}^{++}\big)\bar{{\mathcal{Z}}}^{+}=0\,.
\end{array}
\ee

One of the natural choices of the gauge-invariant action
for nonlinear indecomposable supermultiplet ${\bf (1,4,3){\supset\hspace{-1.1em}+}(4,4,0)}$ multiplet we deal with is the following one:
\be\label{gauge_action}
S_{1}\ =\ \frac{1}{4} \int d\zeta_{\rm H}\,f\left(\mathscr{X}\right)\ +\ \frac{is}{2}\int d\zeta_{\rm A}^{--}\,{\mathcal{V}}^{++}\,,
\ee
where $d\zeta_{\rm H}$ and $d\zeta_{\rm A}^{--}$ are measures of integration over superspaces with the coordinates
$\zeta_{\mathrm{H}}=(t, \theta^{\pm}_{\alpha}, u^\pm_i)$ and
$\zeta_{\mathrm{A}}=(t_{\mathrm{A}}, \theta^{+}_{\alpha}, u^\pm_i)$, respectively (see Appendix).
The action \eqref{gauge_action} is completely specified by the single  function $f\left(\mathscr{X}\right)$. The last term in \eqref{gauge_action}
is the Fayet-Iliopoulos term with $s$ a constant.

Using the gauge transformations \eqref{u1-tr-V}
we can impose the Wess-Zumino gauge, in which ${\mathcal{V}}^{++}$ takes the short form:
\be\label{WZgauge}
{\mathcal{V}}^{++}\left(\zeta_{\rm A}\right)=-\,i\, \theta^{+}_{\alpha}\theta^{+\alpha}{A}\left(t_{\rm A}\right).
\ee
In this gauge, the solutions \eqref{GaugeInvConstraints} and \eqref{eq-143-r} for superfields ${\mathcal{Z}}$ and $\mathscr{X}$,
as well as  their supersymmetry transformations, are given by the expressions \eqref{ZbarZ}, \eqref{X-sol} and \eqref{tr_440}, \eqref{tr-143-l},
with the substitution $\dot z^{i}\to\nabla z^{i}$, $\dot{\bar{z}}_{i}\to\nabla \bar{z}_{i}$, where
\be
\nabla z^{i}=\dot{z}^{i}+i\,{A}\,z^{i},\qquad
\nabla \bar{z}_{i}=\dot{\bar{z}}_{i}-i\,{A}\,\bar{z}_{i}
\ee
are $d=1$ covariant derivatives. Note that the covariant derivatives $\nabla z^{i}$, $\nabla \bar{z}_{i}$ also appear
in the residual supersymmetry transformations as a part of the compensating gauge transformations needed to preserve the Wess-Zumino gauge \eqref{WZgauge}.
As a result, supersymmetry transformations have the form \eqref{tr_440}, with the substitutions  $\dot z^{i}\to\nabla z^{i}$, $\dot{\bar{z}}_{i}\to\nabla \bar{z}_{i}$.

In the Wess-Zumino gauge, the Lagrangian in the invariant action \eqref{gauge_action}, $\displaystyle S_{1}:=\int dt\,{\cal L}_{1}$,
has the following component form
\bea
{\cal L}_{1}&=&
\left(\frac{\dot{x}^2}{2}+\frac{i}{2}\,\psi^{i\alpha}\dot{\psi}_{i\alpha}
-\frac{N^{\alpha\beta}N_{\alpha\beta}}{4}\right)f^{\prime\prime}(x)-
\frac{1}{4}\,N^{\alpha\beta}\psi^i_{\alpha}\psi_{i\beta}\,f^{\prime\prime\prime}(x)-
\frac{1}{24}\,\psi^{i}_{\alpha}\psi_{i\beta}\psi^{j\alpha}\psi^{\beta}_{j}\,
f^{(IV)}(x)\nonumber\\ [5pt]
&&-\,\frac{\varkappa}{2}\Big[i\left(\bar{z}_{i}\nabla z^{i}-z^{i}\nabla\bar{z}_{i}\right)-\xi^{\alpha}\bar{\xi}_{\alpha}\Big]
f^{\prime}\left(x\right)+\frac{\varkappa}{2}\,\psi^{i\alpha}\left(z_{i}\bar{\xi}_{\alpha} + \bar{z}_{i}\xi_{\alpha}\right)f^{\prime\prime}(x)\nonumber\\ [5pt]
&&-\,\frac{\varkappa^2}{8}\left(z^{i}\bar{z}_{i}\right)^2f^{\prime\prime}(x)+
\frac{\varkappa}{4}\,z^{(i}\bar{z}^{j)}\psi_{i}^{\alpha}\psi_{j\alpha}\,
f^{\prime\prime\prime}(x) - s\,{A}\,,\label{L-X}
\eea
where the prime denotes the derivative with respect to the field argument, $f^{\prime}(x)=\partial f(x)/\partial x$, etc.
The Lagrangian \eqref{L-X} is invariant, up to a total derivative, with respect to the residual local transformations,
\be\lb{tr-zA}
z^{\prime}{}^{i} =e^{i\lambda}\,z^{i}\,,\qquad
\bar{z}_{i}^{\prime}=e^{-i\lambda}\,\bar{z}_{i}\,,\qquad
{A}^{\prime} = A - \dot{\lambda}\,,
\ee
\be\lb{tr-xi}
\xi^{\prime\alpha}=e^{i\lambda}\,\xi^{\alpha}\,,\qquad
\bar{\xi}_{\alpha}^{\prime}=e^{-i\lambda}\,\bar{\xi}_{\alpha}
\,,
\ee
where $\lambda= \lambda\left(t\right)$ is $d=1$ gauge parameter.

Eliminating the auxiliary fields $N^{\alpha\beta}$, $\xi^{\alpha}$ and $\bar{\xi}_{\alpha}$ in the Lagrangian \eqref{L-X},
we obtain the on-shell Lagrangian:
\bea
{\cal L}_{1}&=&\left[\frac{\dot{x}^2}{2}+\frac{i}{2}\,\psi^{i\alpha}\dot{\psi}_{i\alpha}-
\frac{\varkappa^2}{8}\left(z^{i}\bar{z}_{i}\right)^2\right]f^{\prime\prime}(x)
-\frac{i}{2}\,\varkappa\left(\bar{z}_{i}\dot{z}^{i}-z^{i}\dot{\bar{z}}_{i}\right)f^{\prime}\left(x\right)
 \nonumber \\ [4pt]
&&{} + \varkappa\,z^{(i}\bar{z}^{j)}\psi_{i}^{\alpha}\psi_{j\alpha}\left[\frac{f^{\prime\prime\prime}(x)}{4}-\frac{\big(f^{\prime\prime}(x)\big)^2} {2\,f^{\prime}\left(x\right)}\right]-
\frac{1}{24}\,\psi^{i}_{\alpha}\psi_{i\beta}\psi^{j\alpha}\psi^{\beta}_{j}\left[f^{(IV)}(x)- \frac{3\,\big(f^{\prime\prime\prime}(x)\big)^2}{2\,f^{\prime\prime}(x)}\right] \nonumber \\ [4pt] &&{} +
{A}\Big[\varkappa\,z^{i}\bar{z}_{i}\,f^{\prime}\left(x\right)-s\Big]\,. \label{n1L} \eea The on-shell supersymmetry transformations read: \bea &&\delta x = \epsilon_{i\alpha}\psi^{i\alpha},\qquad \delta \psi^{i\alpha} =
i\,\epsilon^{i\alpha}\dot{x}-\frac{f^{\prime\prime\prime}(x)}{2\,f^{\prime\prime}(x)}\, \epsilon^{i}_{\beta}\psi^{j\alpha}\psi^{\beta}_{j}+\varkappa\,\epsilon^{\alpha}_{j}z^{(i}\bar{z}^{j)},\nn &&\delta
z^{i}=-\,\frac{f^{\prime\prime}(x)}{f^{\prime}\left(x\right)}\,\epsilon^{i\alpha}\psi_{j\alpha}z^{j},\qquad \delta
\bar{z}_{i}=-\,\frac{f^{\prime\prime}(x)}{f^{\prime}\left(x\right)}\,\epsilon_{i\alpha}\psi^{j\alpha}\bar{z}_{j}\,.
\eea
The Lagrangian \eqref{n1L} respects U$(1)$ gauge symmetry \eqref{tr-zA}. From the form of the
Lagrangian \eqref{n1L} it follows  that the generator of this symmetry is the first-class constraint
\be
\varkappa\,z^{i}\bar{z}_{i}\,f^{\prime}\left(x\right)-s\approx 0\,, \lb{ConstrA}
\ee
for which the $d=1$ field $A(t)$ is the Lagrange multiplier.

The peculiar feature of the Lagrangians \eqref{L-X}, \eqref{n1L} is the presence of the kinetic term of the first order in the time derivative,
\be
    -\,\frac{i}{2}\,\varkappa\left(\bar{z}_{i}\nabla z^{i}-z^{i}\nabla\bar{z}_{i}\right)
f^{\prime}\left(x\right),\label{zdotz}
\ee
that ensures a semi-dynamical interpretation of the multiplet ${\bf (4,4,0)}$.
Analogous WZ terms were also found in the invariant Lagrangians of ${\cal N}=8$ indecomposable multiplets \cite{IS2512}.
At the same time, this peculiarity is absent for the indecomposable multiplet ${\bf (1,4,3){\supset\hspace{-1.1em}+}(3,4,1)}$ \cite{SS2410}. There,  no such terms
of the first order appear from the deformed Lagrangian of the multiplet ${\bf (1,4,3)}$ and,
for the semi-dynamical interpretation of the multiplet ${\bf (3,4,1)}$, one should add
an independent WZ type $d=1$ action for the latter multiplet, with a new constant strength.

In the case of the multiplet ${\bf (1,4,3){\supset\hspace{-1.1em}+}(4,4,0)}$
such a WZ term \eqref{zdotz} appears in the intrinsic minimal way from the deformed ${\bf (1,4,3)}$ action, with the strength proportional to the deformation parameter $\varkappa$.
In principle, one could also extend the action \eqref{gauge_action} by an extra explicit WZ action, with
\be
    S_{\rm WZ} \sim \int d\zeta_{\rm A}^{--}\,c\,\mathcal{Z}^{+}\bar{\mathcal{Z}}^{+}= \int dt\,{\cal L}_{\rm WZ}\,,
    \qquad {\cal L}_{\rm WZ}=-\,\frac{c}{2}\left[i\left(\bar{z}_{i}\nabla z^{i}-z^{i}\nabla\bar{z}_{i}\right)-\xi^{\alpha}\bar{\xi}_{\alpha}\right]. \label{ExtraWZ}
\ee
as the simplest option. In this non-minimal case the new independent WZ constant strength would appear and passing
to the component Lagrangian would lead to a shift $f^{\prime}\left(x\right)\rightarrow f^{\prime}\left(x\right)+c/\varkappa$ in \eqref{L-X}.

The on-shell Lagrangians similar to \eqref{n1L}  appeared earlier also in other approaches, with different origins of the constant $d=1$ WZ strength \cite{BK0905,FIL0905,KL0906,FIL0912}.
All these are related to \eqref{n1L} by some duality transformation. For instance, after the duality transformations of the component fields as \cite{SS2410},
\bea
    y=f^{\prime}\left(x\right),\quad \dot{x}\left(t\right)=x^{\prime}\left(y\right)\dot{y}\left(t\right), \quad \tilde{g}\left(y\right)=x^{\prime}\left(y\right)= \frac{1}{y^{\prime}\left(x\right)}=\frac{1}{f^{\prime\prime}\left(x\right)},\quad \psi^{i\alpha}=\eta^{i\alpha}x^{\prime}\left(y\right),\label{new}
\eea
the Lagrangian \eqref{n1L} proves to be equivalent to that obtained in \cite{BK0905}, with the parameter like $\varkappa$ coming from the interaction term of
two irreducible multiplets ${\bf (1,4,3)}$ and ${\bf (4,4,0)}$. At the off-shell superfield level this duality is seemingly associated with the one  between two
descriptions of the multiplet ${\bf (1,4,3)}$ \cite{IKP1991}. Anyway, the Lagrangians \eqref{L-X}, \eqref{n1L}  with the new mechanism of appearance of $d=1$ WZ term,
were never presented before.

Taking into account the superconformal transformations \eqref{conf},  those of the integration measures,
\be
\delta^{\prime}\left(d\zeta_{\rm H}\right) =
2i\left[\left(1+\alpha\right)\eta^{+\alpha}\theta^{-}_{\alpha}-\left(1-\alpha\right)\eta^{-\alpha}\theta^{+}_{\alpha}\right]d\zeta_{\rm H}\,, \quad
\delta^{\prime}\left(d\zeta_{\rm A}^{--}\right) = 0\,,
\ee
and of the gauge superfield,
$\delta^{\prime}{\mathcal{V}}^{++} = 0$, one can find ${\rm D}(2,1;\alpha)$ superconformally invariant version of the general action \eqref{gauge_action}:
\be
\begin{array}{rcl}
f(\mathscr{X}) &\sim& \ \mathscr{X}^{\,1/(1+\alpha)}\quad {\rm for}\quad \alpha \neq 0\,,\\ [7pt]
f(\mathscr{X}) &\sim& \ \mathscr{X} \log{\mathscr{X}}\ \, \quad {\rm for}\quad \alpha = 0\,.
\end{array}
\ee
It coincides with $\mathcal{N}=4$ superconformal Lagrangians of refs. \cite{BK0905,FIL0905,KL0906,FIL0912}, up to the above-mentioned duality transformations \eqref{new}.
Note that the non-minimal extra term \eqref{ExtraWZ} respects the conformal supersymmetry ${\rm D}(2,1;\alpha)$ only for the choice $\alpha = 0$\,.

 The derivation of one-particle systems  with $\mathcal{N}=4$ superconformal symmetry
in the framework of the indecomposable  supermultiplet suggests, as the natural next step,  the construction of
similar multiparticle matrix systems with $\mathcal{N}=4$ supersymmetry,
in which the parameter $\varkappa$ entering the defining constraints of the relevant indecomposable supermultiplets will play the role of a deformation parameter.
Such a construction should yield generalizations of the matrix superfield systems explored in \cite{FIL0812}.

\section{Indecomposable matrix multiplet (1,4,3)$\,{\supset\hspace{-0.95em}+}\,$(4,4,0)}

Here we consider a matrix generalization of the system described by eqs. \eqref{eq-143-r}.

Instead of scalar field $\mathscr{X}$, we will now deal with the $n\times n$ matrix superfield
\be
\mathscr{X}=\big(\mathscr{X}_b{}^c\big),\qquad b,c=1,\ldots,n\,,
\ee
which is Hermitian with respect to the ``tilde'' conjugation, $\widetilde{\mathscr{X}}=\mathscr{X}$.

The semi-dynamical degrees of freedom of the matrix model will be described by superfields
$\mathcal{Z}^{+}$ and $\bar{\mathcal{Z}}^{+}$, which form a column and a row of the rank $n$:
\be
\mathcal{Z}^{+}=\big(\mathcal{Z}^{+}_b\big),\quad
\bar{\mathcal{Z}}^{+}=\big(\bar{\mathcal{Z}}^{+}{}^b\big)\,, \quad \bar{\mathcal{Z}}^{+}{}^b=\widetilde{\mathcal{Z}^{+}_b}\,.
\ee

Considering the $n\times n$ matrix system requires a wider group to gauge,
namely, the group $\mathrm{U}(n)$.
For this reason, and by analogy with the one-particle system,
we will use the matrix analytic gauge superfield $\mathcal{V}^{++}$:
\be
\mathcal{V}^{++}=\big( \mathcal{V}^{++}\!{}_b{}^c\big),\qquad  \widetilde{\mathcal{V}^{++}}=\mathcal{V}^{++}\,.
\ee
All entries of the matrix superfield $\mathcal{V}^{++}$ are analytic,  $\mathcal{V}^{++}\!{}_b{}^c=\mathcal{V}^{++}\!{}_b{}^c(\zeta_{\mathrm{A}})$.
The matrix generalization of basic equations \eqref{eq-143-r} and \eqref{GaugeInvConstraints} is:
\be\label{Meq-143-r}
D^{+}_{\alpha}D^{+\alpha}\mathscr{X}_b{}^c = 2\varkappa\,\mathcal{Z}_b^{+}\bar{\mathcal{Z}}^{+ c}\,,\qquad D^{++}\mathscr{X}_b{}^c+i\big[\mathcal{V}^{++},\mathscr{X}\big]{}_b{}^c=0
\ee
\be\label{MGaugeInvConstraints}
\begin{array}{l}
D^{+}_{\alpha}\mathcal{Z}_b^{+}=0\,,\qquad D^{++}\mathcal{Z}_b^{+}+i\,\mathcal{V}^{++}\!{}_b{}^c\mathcal{Z}_c^{+}=0\,,
\\ [7pt]
D^{+}_{\alpha}\bar{\mathcal{Z}}^{+}{}^b=0\,,\qquad D^{++}\bar{\mathcal{Z}}^{+}{}^b
-i\,\bar{\mathcal{Z}}^{+}{}^c\mathcal{V}^{++}\!{}_c{}^b=0\,.
\end{array}
\ee
Eqs. \eqref{Meq-143-r} and \eqref{MGaugeInvConstraints}
are covariant with respect to local ${\rm U}(n)$ transformations:
\be\lb{M-u1-tr}
\begin{array}{c}
\displaystyle
\mathscr{X}^{\prime}=e^{i\lambda}\,\mathscr{X}\,e^{-i\lambda}\,,
\qquad {\mathcal{Z}}^{+}{}^{\prime}=e^{i\lambda}\,{\mathcal{Z}}^{+}\,,\qquad
\displaystyle
\bar{{\mathcal{Z}}}^{+}{}^{\prime}=\bar{{\mathcal{Z}}}^{+}\,e^{-i\lambda}\,,
\\ [7pt]
\mathcal{V}^{++}{}^\prime =e^{i\lambda}\left(\mathcal{V}^{++}-i\,D^{++}\right)e^{-i\lambda}\,,
\end{array}
\ee
where $\lambda=\big( \lambda_b{}^c\big)$ is $n\times n$  matrix, all components of which are analytical superfields,
$\lambda_b{}^c=\lambda_b{}^c (\zeta_{\rm A} )$.
The transformations \eqref{M-u1-tr} provide a matrix generalization of the transformations \eqref{u1-tr} and \eqref{u1-tr-V} of the one-particle case.

As a natural choice of the action for the matrix supermultiplet
${\bf (1,4,3){\supset\hspace{-1.1em}+}(4,4,0)}$ we consider the sum of the gauge-invariant action for $\mathscr{X}$
and the Fayet-Iliopoulos action for U$(1)$ component of the gauge superfield $\mathcal{V}^{++}$.
Due to the indecomposable nature of the underlying supermultiplet, such a structure will result in the presence
of all the component fields in the Lagrangian, including those of the superfields ${\mathcal{Z}}^{+}$ and $\bar{{\mathcal{Z}}}^{+}$.

\section{Multiparticle model I: rational case}\label{Sec4}

As the first example, we consider the model where the matrix superfield $\mathscr{X}$ appears quadratically in the superfield Lagrangian. In this case, the invariant quantities are ${\rm Tr}\left(\mathscr{X}^2\right)$ and $\left({\rm Tr}\mathscr{X}\right)^2$.
Thus, we consider a model with the action
\be\lb{L-X-r}
S_{\rm rat.}\ = \
\frac{1}{4} \int d\zeta_{\rm H}\,\Big[{\rm Tr}\big(\mathscr{X}^2\big) + g\big({\rm Tr}\,\mathscr{X}\big)^2\Big]
\ +\ \frac{is}{2}\int d\zeta_{\rm A}^{--}\,{\rm Tr}\,{\cal V}^{++}\,,
\ee
where $s$ and $g$ are real parameters.
As before, the constant $s$ specifies the Fayet-Iliopoulos term of the Lagrangian.
The constant $g$ determines the contribution of the trace part of the superfield  $\mathscr{X}$.
As will be seen below, for a certain value of $g$ the variables corresponding to the center of mass can be removed from consideration.

Although the superfield matrix system \eqref{L-X-r} is relatively simple,
its consideration exhibits most characteristic features of models with matrix indecomposable supermultiplets.

The first step in analyzing the model described by action \eqref{L-X-r} is to properly fix the gauge symmetries.
In supergauge theories, such fixing is the Wess-Zumino gauge,
\be\lb{WZ-m}
\mathcal{V}^{++}\!{}_b{}^c=-\,i\,\theta^{+}_{\alpha}\theta^{+\alpha}{A}_{b}{}^{c}\,,
\ee
where $A(t)=\big({A}_{b}{}^{c}(t) \big)$ is $d=1$ gauge field taking values
in the $u(n)$ algebra.

In the gauge \eqref{WZ-m}, the solution of the constraints
has the form:
\be\lb{sol-Z-m}
\mathcal{Z}^{+}_{b}=Z^{+}_{b}-\theta^{+}_{\alpha}\,\Xi^{\alpha}_{b} + i\,\theta^{+}_{\alpha}\theta^{+\alpha}\,\nabla Z^{-}_{b}\,,\qquad
\bar{\mathcal{Z}}^{+}{}^{b}=\bar{Z}^{+b}-\theta^{+\alpha}\,\bar{\Xi}^{b}_{\alpha}
+i\,\theta^{+}_{\alpha}\theta^{+\alpha}\,\nabla {\bar{Z}}^{-b},
\ee
where $Z_b^{\pm}=Z_b^{i}u_i^\pm$, $\bar Z_b^{\pm}=\bar{Z}^b_{i}u^{\pm i}$ and
\be
\nabla\! Z_b^{i}=\dot{Z}_b^{i}+i\,{A}_b{}^c\,Z_c^{i},\qquad
\nabla\!\bar{Z}^b_{i}=\dot{\bar{Z}}^b_{i}-i\,\bar{Z}^c_{i}\,{A}_c{}^b
\ee
are covariant derivatives of the $d=1$ bosonic fields
$Z^{i}_b$, $\bar{Z}^b_{i}=\big(\overline{Z_b^{i}}\big)$. The fields $\Xi_b^\alpha$, $\bar{\Xi}^b_{\alpha}=\overline{\big(\Xi_b^{\alpha}\big)}$ are fermionic.

Proceeding from the solutions \eqref{sol-Z-m}, we find the solution
of eq. \eqref{Meq-143-r}:
\bea
\mathscr{X}_{b}{}^{c}&=&X_{b}{}^{c}-\theta^{-}_{\alpha}\Psi^{i\alpha}{}_{b}{}^{c}u^{+}_i+
\theta^{+}_{\alpha}\Psi^{i\alpha}{}_{b}{}^{c}u^{-}_i
+\theta^{-}_{(\alpha}\theta^{+}_{\beta)} N^{\alpha\beta}{}_{b}{}^{c}
+i\,\theta^{-}_{\alpha}\theta^{+\alpha}\nabla\! X_{b}{}^{c}
-i\,\theta^{+}_{\alpha}\theta^{+\alpha}\theta^{-}_{\beta}\nabla{\Psi}^{i\beta}{}_{b}{}^{c} u^{-}_i \nonumber \\ [5pt]
&&-\,\frac{\varkappa}{2}\Big(\theta^{+}_{\alpha}\theta^{+\alpha}Z^{-}_{b}\bar{Z}^{-c}
+\theta^{-}_{\alpha}\theta^{-\alpha}Z^{+}_{b}\bar{Z}^{+c}
-\theta^{+}_{\alpha}\theta^{-\alpha}Z^{+}_{b}\bar{Z}^{-c}
-\theta^{+}_{\alpha}\theta^{-\alpha}Z^{-}_{b}\bar{Z}^{+c}\Big)
\nonumber \\ [5pt]
&&-\,\frac{\varkappa}{2}\,\theta^{-}_{\alpha}\theta^{-\alpha} \theta^{+}_{\beta}\Big(Z^{+}_{b}\,\bar{\Xi}^{\beta c}-\Xi^{\beta}_{b}\,\bar{Z}^{+c}\Big)-\frac{\varkappa}{2}\,\theta^{+}_{\alpha}\theta^{+\alpha} \theta^{-}_{\beta}\Big(Z^{-}_{b}\,\bar{\Xi}^{\beta c}-\Xi^{\beta}_{b}\,\bar{Z}^{-c}\Big)
\nonumber \\ [5pt]
&&-\, \frac{\varkappa}{4}\,\theta^{+}_{\alpha}\theta^{+\alpha}\theta^{-}_{\beta}\theta^{-\beta}
\Bigg[2i\Big(\nabla\!{Z}^{-}_{b}\,\bar{Z}^{+c}
+Z^{+}_{b}\,\nabla\!{\bar{Z}}^{-c}\Big)
-\Xi^{\gamma}_{b}\,\bar{\Xi}_{\gamma}^{c}\Bigg]. \lb{sol-X-m}
\eea
Here, the $d=1$ gauge-covariant derivatives are defined as
\be\lb{hat-X-Psi-def}
\nabla\! X_{b}{}^{c} = \dot{X}_{b}{}^{c} + i\big[A,X\big]_{b}{}^{c},\qquad
\nabla \Psi^{i\alpha}{}_{b}{}^{c} = \dot{\Psi}^{i\alpha}{}_{b}{}^{c}+i\big[A,\Psi^{i\alpha}\big]_{b}{}^{c}\,.
\ee
The bosonic matrix fields $X=\big(X_{b}{}^{c}\big)$,
$N^{\alpha\beta}=\big(N^{\alpha\beta}{}_{b}{}^{c}\big)$
and the fermionic once $\Psi^{i\alpha}=\big(\Psi^{i\alpha}{}_{b}{}^{c}\big)$
are subject to the Hermitian conditions
$\big(X\big)^\dagger=X$,
$\big(\Psi^{i\alpha}\big)^\dagger=\Psi_{i\alpha}$,
$\big(N^{\alpha\beta}\big)^\dagger=-\,N_{\alpha\beta}$.

Substituting \eqref{sol-X-m} and \eqref{WZ-m} into the superfield action \eqref{L-X-r} and integrating there over Grassmann and harmonic variables,
we obtain the $d=1$ action
$\displaystyle S_{\rm rat.}=\int dt\,{\cal L}_{\rm rat.}$,
with
\bea
{\cal L}_{\rm rat.}&=&
{\rm Tr}\,\bigg\lbrace\,\frac{1}{2}\,\nabla\!{X}\,\nabla\! X
+\frac{i}{2}\,\Psi^{i\alpha}\,\nabla \Psi_{i\alpha}
-\frac{1}{4}\,N^{\alpha\beta}N_{\alpha\beta}-s\,{A}\,\bigg\rbrace
\nonumber \\ [5pt]
&& {}+g\,\bigg\lbrace\,\frac{1}{2}\,\dot{X}_b{}^b\,\dot X_c{}^c+\frac{i}{2}\,\Psi^{i\alpha}{}_{b}{}^{b}{}_{}\,\dot{\Psi}_{i\alpha c}{}^{c}-\frac{1}{4}\,N^{\alpha\beta}{}_{b}{}^{b}N_{\alpha\beta c}{}^{c}\bigg\rbrace
\nonumber \\ [5pt]
&& {}-\frac{\varkappa}{2}\, \hat X_{b}{}^{c} \Big[i\big(\nabla\! Z_c^{i}\,\bar{Z}^b_{i}
-Z_c^{i}\,\nabla\! \bar{Z}^b_{i}\big)-\Xi_c^{\alpha}\bar{\Xi}^b_{\alpha}\Big]
\nonumber \\ [5pt]
&& {}+\,\frac{\varkappa}{2}\,\hat\Psi^{i\alpha}{}_{b}{}^{c} \Big(Z_{ic}\bar{\Xi}^b_{\alpha} + \Xi_{\alpha c}\bar{Z}^b_{i}\Big) +\frac{\varkappa^2}{4}\,\Big[Z_b^{(i}\bar{Z}^{j)c}\,Z_{ic}\bar{Z}^b_{j}
+g\,Z_b^{(i}\bar{Z}^{j)b}\,Z_{ic}\bar{Z}^c_{j}\Big], \lb{La-X-m}
\eea
where
\be\lb{def-hat-X-Psi}
\hat X_{b}{}^{c}  =  X_{b}{}^{c}+g\,\delta_{b}^{c}\,X_{d}{}^{d}\,,\qquad
\hat\Psi^{i\alpha}{}_{b}{}^{c}  =  \Psi^{i\alpha}{}_{b}{}^{c}+g\,\delta_{b}^{c}\,\Psi^{i\alpha}{}_{d}{}^{d}\,.
\ee
The Lagrangian \eqref{La-X-m}, apart from the ${\rm U}(n)$ gauge symmetry, exhibits also superconformal
symmetry ${\rm D}(2,1;-1/2) = {\rm OSp}(4|2)$, as we will show soon.

In what follows, we will consider the cases in which the condition $\det\hat X\neq 0$ is satisfied.
Eliminating the auxiliary fields by their equations of motion, we find the on-shell Lagrangian in the form
\bea
{\cal L}^{\rm on-shell}_{\rm rat.}&=&\frac{1}{2}\left(\nabla{X}_{b}{}^{c}\,\nabla{X}_{c}{}^{b}+g\,\dot{X}_b{}^b\,\dot X_c{}^c\right)+\frac{i}{2}\left(\Psi^{i\alpha}{}_{b}{}^{c}\,\nabla{\Psi}_{i\alpha}{}_{c}{}^{b}+g\,
\Psi^{i\alpha}{}_{b}{}^{b}{}_{}\,\dot{\Psi}_{i\alpha c}{}^{c}\right)\nn [5pt]
&&-\,\frac{i}{2}\,\varkappa\hat X_{b}{}^{c}\Big(\nabla{Z}^{i}_{c}\,\bar{Z}^{b}_{i}-Z^{i}_{c}\,\nabla{\bar{Z}}^{b}_{i}\Big)
+\frac{\varkappa^2}{4}\,\Big[Z_b^{(i}\bar{Z}^{j)c}\,Z_{ic}\bar{Z}^b_{j}
+g\,Z_b^{(i}\bar{Z}^{j)b}\,Z_{ic}\bar{Z}^c_{j}\Big]\nn [5pt]
&&+\,\frac{\varkappa}{2}\,Z^{i}_{a}\bar{Z}^{jb}\hat\Psi_{j\alpha}{}_{b}{}^{c}
\big(\hat X{}^{-1}\big)_{c}{}^{e}\hat\Psi^{\alpha}_{i}{}_{e}{}^{a}
-s\,A_{a}{}^{a}\,, \label{onshell-L}
\eea
where $\big(\hat X{}^{-1}\big)_{c}{}^{b}$ is an inverse to $\hat X_{b}{}^{c}$:
$\hat X_{b}{}^{c}\big(\hat X{}^{-1}\big)_{c}{}^{a}=\delta_b^a$.
In  the system \eqref{onshell-L}, the conjugate momenta of all the component fields have the form:
\be\lb{m-PX}
P_b{}^c = \nabla{X}_{b}{}^{c}+g\,\delta_{b}^{c}\,\dot{X}_{d}{}^{d}\,,
\ee
\be\lb{m-PZ}
P_{\Psi}\,{}^{i\alpha}{}_{a}{}^{b}=
{}\frac{i}{2}\,
\hat\Psi^{i\alpha}{}_{a}{}^{b}\,,
\qquad\quad
P_Z\,{}_i^b = {}-\frac{i}{2}\,\varkappa\, \bar{Z}^c_{i}
\hat X_{c}{}^{b}\,,\qquad
P_{\bar Z}\,{}_b^i= \frac{i}{2}\,\varkappa\,
\hat X_{b}{}^{c}\,,
\ee
\be\lb{m-PA}
P_A\,{}_b{}^c = 0\,.
\ee
These expressions yield the primary constraints to be discussed below.

The on-shell supersymmetry transformations of the matrix component fields
in the Lagrangian \eqref{onshell-L} are given by
\be\label{onshell-tr}
\begin{array}{c}
\delta X_{a}{}^{b} = \epsilon_{i\alpha}\Psi^{i\alpha}{}_{a}{}^{b}\,,\qquad
\delta \Psi^{i\alpha}{}_{a}{}^{b} = i\,\epsilon^{i\alpha}\nabla X_{a}{}^{b}+\varkappa\,\epsilon^{\alpha}_{j}Z^{(i}_{a}\bar{Z}^{j)b}\,,\\ [6pt]
\delta Z^{i}_{a}=-\,\epsilon^{i\alpha}\big(\hat X{}^{-1}\big){}_{a}{}^{b}
\hat\Psi_{j\alpha}{}_{b}{}^{c} Z^{j}_{c}\,,\qquad
\delta \bar{Z}^{a}_{i}=-\,\epsilon_{i\alpha}\,\bar{Z}^{c}_{j}
\hat\Psi^{j\alpha}{}_{c}{}^{b}
\big(\hat X{}^{-1}\big){}_{b}{}^{a}\,.
\end{array}
\ee
The conserved supercharges $Q^{i\alpha}$ are derived in the standard way using the variations \eqref{onshell-tr} of the Lagrangian \eqref{onshell-L}
and have the form:
\be\label{Q}
Q^{i\alpha} = i\,P_{a}{}^{b}\,\Psi^{i\alpha}{}_{b}{}^{a}-\varkappa\,
Z^{(i}_{a}\bar{Z}^{j)b}\hat\Psi^{\alpha}_{j}{}_{b}{}^{a}\,.
\ee

Below we will separately consider the cases $\displaystyle g\,{\neq}-1/n$ and $\displaystyle g\,{=}-1/n$ ,
since our system has different numbers of constraints and final degrees of freedom for these two options.

\subsection{The case with center of mass, $\displaystyle g\,{\neq}-\frac{1}{n}$}

\subsubsection{Supersymmetric matrix system}\label{matrix-system}

The system with Lagrangian
\eqref{m-PX} is described by
the Hamiltonian
\bea
H_{\rm rat.} &=& H- {\rm Tr}\big(A\,G\big)\,, \label{H-rat-t} \\ [5pt]
H &=& {}\frac{1}{2}\,\Big(P_{b}{}^{c}P_{c}{}^{b}- \frac{g}{1+ng}\,P_{b}{}^{b}P_{c}{}^{c}\Big)-
\frac{\varkappa^2}{4}\,\Big[Z_b^{(i}\bar{Z}^{j)c}\,Z_{ic}\bar{Z}^b_{j}
+g\,Z_b^{(i}\bar{Z}^{j)b}\,Z_{ic}\bar{Z}^c_{j}\Big]\nn [5pt]
&&{}-\frac{\varkappa}{2}\,Z^{i}_{a}\bar{Z}^{jb}\hat\Psi_{j\alpha}{}_{b}{}^{c}\big(\hat X{}^{-1}\big){}_{c}{}^{e}
\hat\Psi^{\alpha}_{i}{}_{e}{}^{a}  \label{H-rat}
\eea
and by the primary constraints
\bea
&&C\,{}^{i\alpha}{}_{a}{}^{b}:=
P_{\Psi}\,{}^{i\alpha}{}_{a}{}^{b}-\frac{i}{2}\hat\Psi^{i\alpha}{}_{a}{}^{b}\approx 0\,,
\label{primary-Psi}\\ [4pt]
&&C\,{}_i^b:=P_Z\,{}_i^b+\frac{i}{2}\,\varkappa\, \bar{Z}^c_{i}
\hat X_{c}{}^{b} \approx 0\,,\qquad
\bar C\,{}_b^i:=P_{\bar Z}\,{}_b^i - \frac{i}{2}\,\varkappa\,
\hat X_{b}{}^{c} Z_c^{i}\approx 0,\label{primary-Z}\\ [5pt]
&&P_A\,{}_b{}^c\approx  0\,,\label{primary-A}
\eea
where
\be\label{Gab}
G_{a}{}^{b} = i\left(X_{a}{}^{c}P_{c}{}^{b}-P_{a}{}^{c}X_{c}{}^{b}\right)+
\frac{\varkappa}{2}\left(\hat X_{a}{}^{c} Z^{i}_{c}\bar{Z}^{b}_{i}+Z^{i}_{a}\bar{Z}^{c}_{i}\hat X_{c}{}^{b}\right)
+\,\Psi^{i\alpha}{}_{a}{}^{c}\Psi_{i\alpha}{}_{c}{}^{b}-s\,\delta_{a}^{b}\,.
\ee
The constraints \eqref{primary-Psi}, \eqref{primary-Z} and \eqref{primary-A}
are a direct consequence of expressions \eqref{m-PZ} and \eqref{m-PA} for the conjugated momenta.

As a condition of preserving the constraints \eqref{primary-A} we obtain the secondary constraints
\be
G_a{}^b\approx 0.\label{Gab-0}
\ee
The variables $A_{a}{}^{b}$ serve as Lagrangian multipliers for these constraints.

The non-vanishing Poisson brackets of the constraints \eqref{primary-Psi} and \eqref{primary-Z}
are as follows:
\be\lb{C-PB}
\left\lbrace C_{i\alpha}{}_{a}{}^{b},C^{j\beta}{}_{c}{}^{d}\right\rbrace = {}-i\,\delta_{i}^{j}\,\delta^{\beta}_{\alpha}\left(\delta_{a}^{d}\,\delta_{c}^{b}+
g\,\delta_{a}^{b}\,\delta_{c}^{d}\right)\,,
\qquad
\left\lbrace \bar C^{i}_{a},C^{b}_{j}\right\rbrace  =  {}-i \varkappa \,\delta_{j}^{i}\hat X_{a}{}^{b}\,.
\ee
Since for $\displaystyle g\,{\neq}-1/n$  the right-hand sides in \eqref{C-PB} are non-singular matrices,
the constraints \eqref{primary-Psi} and \eqref{primary-Z} are second class.
To account for them, we introduce the corresponding Dirac brackets.
The constraints \eqref{primary-Psi} and \eqref{primary-Z} are then assumed to be satisfied in the strong sense.
The Dirac brackets for the phase variables are as follows:
\bea
&&\left\lbrace X_{a}{}^{b},P_{c}{}^{d}\right\rbrace^*=\delta_{a}^{d}\,\delta_{c}^{b}\,,\qquad \left\lbrace Z^{i}_{a},\bar{Z}^{b}_{j}\right\rbrace^* = \frac{i}{\varkappa}\,\delta_{j}^{i}\big(\hat X{}^{-1}\big){}_{a}{}^{b},\lb{DB-b}\\
&&\left\lbrace P_{a}{}^{b},P_{c}{}^{d}\right\rbrace^*=\frac{i}{4}\,\varkappa\,\bigg\lbrace Z^{i}_{c}\bar{Z}^{b}_{i}\big(\hat X{}^{-1}\big){}_{a}{}^{d} -Z^{i}_{a}\bar{Z}^{d}_{i}\big(\hat X{}^{-1}\big){}_{c}{}^{b} +g\,\delta_{a}^{b}\left[Z^{i}_{c}\bar{Z}^{e}_{i}
\big(\hat X{}^{-1}\big){}_{e}{}^{d} -Z^{i}_{e}\bar{Z}^{d}_{i}
\big(\hat X{}^{-1}\big){}_{c}{}^{e}\right]\nn
&&\qquad\qquad\qquad\quad+\,
g\,\delta_{c}^{d}\left[Z^{i}_{e}\bar{Z}^{b}_{i}
\big(\hat X{}^{-1}\big){}_{a}{}^{e} -Z^{i}_{a}\bar{Z}^{e}_{i}\big(\hat X{}^{-1}\big){}_{e}{}^{b}\right]
\bigg\rbrace\,,\lb{DB-PP} \\
&&\left\lbrace P_{a}{}^{b},Z^{i}_{c}\right\rbrace^*
=\frac{1}{2}\left[Z^{i}_{a}\big(\hat X{}^{-1}\big){}_{c}{}^{b}+g\,\delta_{a}^{b}\,Z^{i}_{d}
\big(\hat X{}^{-1}\big){}_{c}{}^{d}\right],\lb{DB-PZ}\\
&&\left\lbrace P_{a}{}^{b},\bar{Z}^{c}_{i}\right\rbrace^*
=\frac{1}{2}\left[\bar{Z}^{b}_{i}\big(\hat X{}^{-1}\big){}_{a}{}^{c}+g\,\delta_{a}^{b}\,\bar{Z}^{d}_{i}\big(\hat X{}^{-1}\big){}_{d}{}^{c}\right],\lb{DB-PZb}\\
&&\left\lbrace \Psi_{i\alpha}{}_{a}{}^{b},\Psi^{j\beta}{}_{c}{}^{d}\right\rbrace^*
=-\,i\,\delta_{i}^{j}\,\delta^{\beta}_{\alpha}\left(\delta_{a}^{d}\,
\delta_{c}^{b}-\frac{g\,\delta_{a}^{b}\,\delta_{c}^{d}}{1+ng}\right).
\label{DB-f}
\eea

It is easy to check that, with respect to the Dirac brackets \eqref{DB-b}-\eqref{DB-f}, the constraints \eqref{Gab}, \eqref{Gab-0} form  $u(n)$ algebra:
\be
\left\lbrace G_{a}{}^{b},G_{c}{}^{d}\right\rbrace^*=
i\left(\delta_{a}^{d}\,G_{c}{}^{b}-\delta_{c}^{b}\,G_{a}{}^{d}\right).
\ee
Therefore, the residual constrains \eqref{Gab-0} are first class.

The matrix superfield model \eqref{L-X-r} is invariant with respect
to the superconformal transformations \eqref{tr-d21a} and \eqref{conf} at $\alpha=-1/2$.
The corresponding Noether supercharges obtained by the standard routine have the form:
\be\label{Q-conf}
{S}^{i\alpha}= i\,\hat X_{a}{}^{b}\,\Psi^{i\alpha}{}_{b}{}^{a}\,.
\ee

The supertranslations \eqref{Q} and the supercharges \eqref{Q-conf} form the odd
sector of the conformal $osp(4|2)$ superalgebra. Their closure,
\be
\left\lbrace Q^{i\alpha}, Q^{j\beta}\right\rbrace^* = -\,2\,\varepsilon^{ij}\,\varepsilon^{\alpha\beta}\,H \,,\qquad \left\lbrace {S}^{i\alpha},
{S}^{j\beta}\right\rbrace^* = -\,2\,\varepsilon^{ij}\,\varepsilon^{\alpha\beta}\,K\,,
\ee
\be
\left\lbrace Q^{i\alpha}, {S}^{j\beta}\right\rbrace^* =
2\,\varepsilon^{ij}\,\varepsilon^{\alpha\beta}\,D-i\,\varepsilon^{ij}\,F^{\alpha\beta}-i\,\varepsilon^{\alpha\beta}\,I^{ij}\,,
\ee
produces the full set of bosonic conserved charges: the generator $H$ defined in \eqref{H-rat} and the remaining bosonic generators
\be\lb{bos-conf1}
D=-\,\frac{1}{2}\,X_{a}{}^{b} P_{b}{}^{a}\,,\qquad
K=\frac{1}{2}\,X_{a}{}^{b} \hat X_{b}{}^{a}\,,
\ee
\be
I^{ij}=\varkappa\,Z^{(i}_{a}\bar{Z}^{j)b}\hat X_{b}{}^{a}
-\frac{1}{2}\,\Psi^{(i}_{\alpha}{}_{a}{}^{b}\,\hat\Psi^{j)\alpha}{}_{b}{}^{a}\,,\qquad
F^{\alpha\beta}=\frac{1}{2}\,\Psi^{i(\alpha}{}_{a}{}^{b}\hat\Psi^{\beta)}_{i}{}_{b}{}^{a}\,.\label{bos-conf}
\ee
The generators $\left(H,D,K\right)$ constitute the conformal algebra $so(2,1)$,
\be
\left\lbrace H \,, K\right\rbrace^*=2i\,D\,,\qquad\left\lbrace D,K\right\rbrace^*=i\,K\,,\qquad\left\lbrace D,H \right\rbrace^*=-\,i\,H \,,\label{conf-alg}
\ee
while $I^{ij}$ and $F^{\alpha\beta}$ correspond to the algebra $su(2)\oplus su(2)$:
\be
\left\lbrace I^{ij},I^{kl}\right\rbrace^*=\varepsilon^{ik}\,I^{jl}+\varepsilon^{jl}\,I^{ik}\,,
\qquad \left\lbrace F^{\alpha\beta},F^{\gamma\delta}\right\rbrace^*=\varepsilon^{\alpha\gamma}\,F^{\beta\delta}+\varepsilon^{\beta\delta}\,F^{\alpha\gamma} \,.
\ee
The Dirac brackets between even charges \eqref{H-rat}, \eqref{bos-conf1}, \eqref{bos-conf} and odd generators
\eqref{Q}, \eqref{Q-conf} are collected as:
\bea
&&\left\lbrace K,Q^{i\alpha}\right\rbrace^*=i\,{S}^{i\alpha},\quad \left\lbrace H \,, {S}^{i\alpha}\right\rbrace^*=-\,i\,Q^{i\alpha},\quad \left\lbrace D,Q^{i\alpha}\right\rbrace^*=-\,\frac{i}{2}\,Q^{i\alpha},\quad \left[D,{S}^{i\alpha}\right]=\frac{i}{2}\,{S}^{i\alpha},\nn
&&\left\lbrace I^{ij},Q^{k\alpha}\right\rbrace^*=-\,\varepsilon^{k(i}\,Q^{j)\alpha},\qquad
\left\lbrace I^{ij},{S}^{k\alpha}\right\rbrace^*=-\,\varepsilon^{k(i}\,{S}^{j)\alpha},\qquad
\nn
&&\left\lbrace F^{\alpha\beta},Q^{i\gamma}\right\rbrace^*=-\,\varepsilon^{\gamma(\alpha}\,Q^{i\beta)},\qquad
\left\lbrace F^{\alpha\beta},{S}^{i\gamma}\right\rbrace^*=-\,\varepsilon^{\gamma(\alpha}\,{S}^{i\beta)}\,.
\label{osp42}
\eea

Thus, the superfield system \eqref{L-X-r}  possesses $\mathcal{N}=4$  superconformal symmetry $osp(4|2)$.

Next, we find out the physical field contents of the bosonic core of the considered system.

\subsubsection{Bosonic core}\label{bos-core}

We introduce the following notation for the entries of matrices $X$ and $P$:
\be\label{matrix-XP}
\begin{array}{ll}
x_a = X_a{}^a \,,\qquad & p_a:= P_a{}^a \qquad \mbox{(no summation over $a$)}\,,\\ [6pt]
x_a{}^b= X_a{}^b \,,\qquad & p_a{}^b:= P_a{}^b  \qquad \mbox{for}\quad a\neq b\,,
\end{array}
\ee
{\it i.e.}, $X_a{}^b =x_a\,\delta_a^b +x_a{}^b$, $P_a{}^b =p_a\,\delta_a^b +p_a{}^b$.
In these notations, the bosonic limit of the constraints \eqref{Gab}, \eqref{Gab-0} takes the form
\be\label{T-nondiag}
G_a{}^b = i(x_a-x_b)p_a{}^b - i(p_a-p_b)x_a{}^b  +i(x_a{}^c p_c{}^b-p_a{}^c x_c{}^b) +
T_a{}^b \approx 0
\ee
for $a\neq b$ and
\be\label{T-diag}
G_a{}^a = i(x_a{}^c p_c{}^a-p_a{}^c x_c{}^a) +
T_a{}^a - s\approx 0\,,  \qquad \mbox{(no summation over $a$)},
\ee
for diagonal elements of the matrix $G$, where
\be\label{T-def}
T_a{}^b = \frac{\varkappa}{2}\left(\hat x_{a}+\hat x_b\right)Z^{i}_{a}\bar{Z}^{b}_{i}
+\frac{\varkappa}{2}\left(x_{a}{}^{c}\,Z^{i}_{c}\bar{Z}^{b}_{i}+
Z^{i}_{a}\bar{Z}^{c}_{i}\,x_{c}{}^{b}\right).
\ee
In \eqref{T-def}, we use the notation
\be\label{hat-x-def}
\hat x_a=x_{a}+ng\,x_0\,,
\ee
where $\hat x_a$ are the diagonal elements of the matrix $\hat X$ defined in \eqref{hat-X-Psi-def}
and
\be\label{x0-def}
x_0=\frac{1}{n}\,\sum\limits_a x_a
\ee
stands for the center-of-mass coordinate.

In the case of the Calogero-like conditions $x_a \neq x_b$, we can
impose the gauge
\be\label{x-fix}
x_a{}^b \approx 0\,, \qquad a\neq b\,,
\ee
related to the constraints (\ref{T-nondiag}). Then we introduce Dirac brackets for the constraints  (\ref{T-nondiag}), (\ref{x-fix}) and
eliminate $x_a{}^b$ by (\ref{x-fix}) and $p_a{}^b$ by (\ref{T-nondiag}).
As a result, we have
\be\label{pab}
p_a{}^b =\frac{i\,T_a{}^b}{x_a - x_b}\,,\qquad \qquad a\neq b\,,
\ee
in which
\be\label{Tab}
T_a{}^b = \frac{\varkappa}{2}
\left(\hat x_{a}+\hat x_b\right) Z^{i}_{a}\bar{Z}^{b}_{i}\,.
\ee
Note that, due to the form of the gauge fixing condition \eqref{x-fix},
the Dirac brackets for the remaining phase variables do not change.

In the gauge (\ref{x-fix}),  the bosonic sector of the Hamiltonian \eqref{H-rat} takes the form:
\bea
\nonumber
H_{\rm rat.} &=& \frac12 \left[\sum_a  (p_a)^2 -\frac{g}{1+ng}\left(\sum_a p_a\right)^2\right]
+\frac12\, \sum_{a\neq b} \frac{T_a{}^b T_b{}^a}{(x_a - x_b)^2} \\[7pt]
&& \displaystyle {}  -
\frac{\varkappa^2}{4}\,\sum_{a, b} \Big[Z_b^{(i}\bar{Z}^{j)c}\,Z_{ic}\bar{Z}^b_{j}
+g\,Z_b^{(i}\bar{Z}^{j)b}\,Z_{ic}\bar{Z}^c_{j}\Big] - \sum_a A_a{}^aG_a{}^a.
\label{Ham-CM}
\eea
The residual diagonal constraints \eqref{T-diag} become
\be\label{T-diag-f}
G_a{}^a \ = \
T_a{}^a -s \ = \ \varkappa\,\hat x_{a}Z^{i}_{a}\bar{Z}^{a}_{i} - s\ \approx\ 0  \qquad \mbox{(no summation over $a$)}.
\ee

To simplify $T_a{}^b$ and the Dirac brackets \eqref{DB-b}-\eqref{DB-PZb}, we pass to the new phase variables.
Instead of $Z_a^{i}$ and $\bar{Z}^a_{i}$, we introduce the variables\footnote{\label{new-phase-footnote}
This choice is prompted by the property  that in the gauge \eqref{x-fix} the constraints \eqref{primary-Z} can be rewritten as
\be
P_Z\,{}_i^a+\frac{i}{2}\,\varkappa\,\hat x_{a} \bar{Z}^a_{i} \approx 0,\qquad
P_{\bar Z}\,{}_a^i - \frac{i}{2}\,\varkappa\,\hat x_{a} Z_a^{i}\approx 0\,.\label{primary-Z-s}
\ee
}
\be\lb{new-Z-m-b}
{\it z}_a^{i}=\sqrt{\varkappa\,\hat x_{a}}\,Z_a^{i}\,,\qquad
\bar{{\it z}}^a_{i}=\sqrt{\varkappa\,\hat x_{a}}\,\bar{Z}^a_{i}
\ee
In addition, instead of the coordinates $x_a$ and momenta $p_a$,
we introduce new coordinates $\hat x_a$ defined in
\eqref{hat-x-def}, as well as the new momenta
\be
\hat p_a=p_a-\frac{g}{1+ng}\,\sum_b p_b\,.\label{new-phase}
\ee
In terms of the new variables, the non-zero Dirac brackets \eqref{DB-b}-\eqref{DB-PZb} become
\be\label{Dir-br-n-n}
\left\{\hat x_a, \hat p_b \right\}^*=\delta_{ab}\,,
\qquad
\left\{{{\it z}}_a^i, \bar {{\it z}}^b_j \right\}^*
= i\,\delta^{i}_j\,\delta^{b}_{a}\,.
\ee
At the same time,
the constraints \eqref{T-diag-f} are rewritten as
\be\label{T-diag-ff}
G_a{}^a =
T_a{}^a - s =  {{\it z}}_a^{i}\bar{{\it z}}^a_{i} - s\approx 0  \qquad \mbox{(no summation over $a$)}\,,
\ee
while the non-diagonal $T_a{}^b$ are expressed as
\bea
    T_a{}^b = \frac{\left(\hat x_{a}+\hat x_{b}\right)z^{i}_{a}\bar{z}^{b}_{i}}
{2\sqrt{\hat x_{a}\hat x_{b}}}\,.
\eea
Let us emphasize that for $g\neq {}-1/n$, the variables $\hat x_{a}$ are independent.

The Hamiltonian  \eqref{Ham-CM}, up to the last terms involving constraints,
takes the form:
\bea
H_{\rm rat.} &=&\displaystyle \frac12\,
\bigg[\sum_a  \hat p_a\hat p_a +g\Big(\sum_a  \hat p_a\Big)^2\bigg]
{}+\frac12\,\sum_{a\neq b}
\frac{J_a{}_{i}{}^j\,J_b{}_{j}{}^i}{(\hat x_a - \hat x_b)^2 }  \nonumber\\ [6pt]
&&{}+
\sum_{a\neq b} \frac{1}{8\,\hat x_a \hat x_b} \Big[(2-g)\,J_a{}_{i}{}^i\,J_b{}_{j}{}^j +2g\,J_a{}_{i}{}^j\,J_b{}_{j}{}^i\Big]{}+
\sum_{a} \frac{1+g}{8(\hat x_a)^2}\, J_a{}_{i}{}^i\,J_a{}_{j}{}^j
\,.
\label{Ham-CM-nnnn-0}
\eea
where
\be\label{S-def-n-0}
J_a{}_{i}{}^j=\bar{{\it z}}^a_{i}{{\it z}}_a^{j} \qquad\quad \mbox{(no summation over $a$)}
\ee
are the generators of $u(2)$ algebra with respect to the Dirac brackets \eqref{Dir-br-n-n} at any index $a$.
We can also rewrite   the conformal generators \eqref{bos-conf1} through new variables:
\be
D=-\,\frac{1}{2}\,\sum_{a}\hat{x}_{a}\hat{p}_{a}\,,\qquad
K=\frac{1}{2}\left[\sum_{a}\left(\hat{x}_{a}\right)^2-\frac{g}{1+ng}\left(\sum_{a}\hat{x}_{a}\right)^2\right].\label{bos-conf1-hat}
\ee
Together with the Hamiltonian \eqref{Ham-CM-nnnn-0}, they form the conformal algebra \eqref{conf-alg}.

The first term in \eqref{Ham-CM-nnnn-0} is not diagonalized. For its diagonalization, we note that, when $g\neq 0$,
the following relations take place:
$$
\sum_a  \hat p_a\hat p_a +g\Big(\sum_a  \hat p_a\Big)^2=\sum_a  \rho_a\rho_a \,,
$$
where
$$
\rho_a=\hat p_a+\beta\sum_b  \hat p_b\,,\qquad \beta=\frac{-1\pm\sqrt{1+ng} }{n} \,.
$$
$$
y_a=\hat x_a+\gamma\sum_b  \hat x_b\,,\qquad
\gamma={}\mp \frac{\beta}{\sqrt{1+ng}}={}-\frac{1}{n}\Big(1\mp\frac{1}{\sqrt{1+ng} } \Big)\,.
$$
In this case,  we learn from \eqref{Dir-br-n-n} that the variables $y_a$ and $\rho_a$ are canonically conjugate:
\be\label{Dir-br-n-n-y}
\left\{y_a, \rho_b \right\}^*=\delta_{ab}\,.
\ee
By passing to the variables $y_a$ and $\rho_a$, we arrive at the bosonic system, which is some deformation of the $n$-particle Calogero spin system.
Thus, the full system we have constructed provides $\mathcal{N}{=}4$ supersymmetrization of this bosonic system, such that the full system  is also ${\rm OSp}(4|2)$ superconformal.
One can check that the conformal generators \eqref{bos-conf1-hat} are also diagonalized after passing to the new variables.

Note that the  resulting bosonic system with the Hamiltonian \eqref{Ham-CM} differs from the standard Calogero system.
Possible implications of this deformed bosonic system (and its possible integrability) are still to be recognized.

\subsection{The system without center of mass, $\displaystyle g\,{=}-\frac{1}{n}$}

The matrices \eqref{def-hat-X-Psi}
at $\displaystyle g\,{=}-\frac{1}{n}$, \textit{i.e.},
\be\lb{tr-hat-X-Psi}
\hat X_{b}{}^{c}  =  X_{b}{}^{c}-\frac{1}{n}\,\delta_{b}^{c}\,X_{d}{}^{d}\,,\qquad
\hat\Psi^{i\alpha}{}_{b}{}^{c}  =  \Psi^{i\alpha}{}_{b}{}^{c}-\frac{1}{n}\,\,\delta_{b}^{c}\,\Psi^{i\alpha}{}_{d}{}^{d}\,,
\ee
are traceless,
\be\lb{tr0-hat-X-Psi}
\hat X_{b}{}^{b}  =  0\,,\qquad
\hat\Psi^{i\alpha}{}_{b}{}^{b}  = 0\,.
\ee
This leads to a number of special features of the model under consideration.

Firstly, as follows from \eqref{m-PX}, in the case $\displaystyle g\,{=}-\frac{1}{n}$
there appears an additional primary constraint:
\be\lb{constr-P0-m-b}
P{}_b{}^b\approx 0.
\ee
It shows that the model describes a system without the center-of-mass coordinate.

In this case the Hamiltonian has the form \eqref{H-rat-t},
where
\be
H = {}\frac{1}{2}\, P_{b}{}^{c}P_{c}{}^{b} -
\frac{\varkappa^2}{4}\,\Big[Z_b^{(i}\bar{Z}^{j)c}\,Z_{ic}\bar{Z}^b_{j}
-\frac1{n}\,Z_b^{(i}\bar{Z}^{j)b}\,Z_{ic}\bar{Z}^c_{j}\Big]
{}-\frac{\varkappa}{2}\,Z^{i}_{a}\bar{Z}^{jb}\hat\Psi_{j\alpha}{}_{b}{}^{c}\big(\hat X{}^{-1}\big){}_{c}{}^{e}
\hat\Psi^{\alpha}_{i}{}_{e}{}^{a}\,.  \label{H-rat0}
\ee
Here
$\hat X_{b}{}^{c}$ and
$\hat\Psi^{i\alpha}{}_{b}{}^{c}$ are defined in \eqref{tr-hat-X-Psi}.
As before, the variables $A{}_{b}{}^{c}$ are Lagrangian multipliers for the constraints
$G_a{}^b\approx 0$ (see \eqref{Gab-0}), where now
\be\label{Gab0}
G_{a}{}^{b} = i\left[X,P\right]_{a}{}^{b}+
\frac{\varkappa}{2}\left(\hat X_{a}{}^{c} Z^{i}_{c}\bar{Z}^{b}_{i}+Z^{i}_{a}\bar{Z}^{c}_{i}\hat X_{c}{}^{b}\right)
+\,\hat \Psi^{i\alpha}{}_{a}{}^{c}\hat \Psi_{i\alpha}{}_{c}{}^{b}-s\,\delta_{a}^{b}\,.
\ee

As before, the primary  constraints are \eqref{primary-Psi} and \eqref{primary-Z}.
We saw that for $\displaystyle g\,{\neq}-\frac{1}{n}$, all of these constraints were second class. In particular,
the fermionic $n^2$ constraints are second class. But this is not true for $\displaystyle g\,{=}-\frac{1}{n}$.
Now, among the fermionic $n^2$ constraints, the constraint
\be\label{primary-Psi-0}
C\,{}^{i\alpha}{}_{b}{}^{b}=
P_{\Psi}\,{}^{i\alpha}{}_{b}{}^{b}\approx 0
\ee
is first class, while the remaining $n^2-1$ fermionic constraints,
\be\label{primary-Psi-tr}
\hat C\,{}^{i\alpha}{}_{a}{}^{b}:=
\hat P_{\Psi}\,{}^{i\alpha}{}_{a}{}^{b}-\frac{i}{2}\hat\Psi^{i\alpha}{}_{a}{}^{b}\approx 0\,,
\ee
are second class, where
$\displaystyle \hat P_{\Psi}\,{}^{i\alpha}{}_{a}{}^{b}=
P_{\Psi}\,{}^{i\alpha}{}_{a}{}^{b}-\frac{1}{n}\,
\delta_a^b P_{\Psi}\,{}^{i\alpha}{}_{c}{}^{c}$. Since for
$\displaystyle g\,{=}-\frac{1}{n}$,
$$
\big\lbrace \hat X_{a}{}^{b},P_{c}{}^{c}\big\rbrace=0\,,
\qquad \big\lbrace \hat\Psi^{i\alpha}{}_{a}{}^{b},P_{\Psi}\,{}^{i\alpha}{}_{c}{}^{c}\big\rbrace=0\,,
$$
the constraints \eqref{constr-P0-m-b} and \eqref{primary-Psi-0} have zero Poisson brackets
with all constraints and with the Hamiltonian \eqref{H-rat0}.

Using the first-class constraint \eqref{primary-Psi-0}, we can impose the gauge-fixing condition
\be\label{con-Psi-ga}
\Psi\,{}^{i\alpha}{}_{b}{}^{b}\approx 0\,.
\ee
After this, all fermionic constraints \eqref{primary-Psi-tr}, \eqref{primary-Psi-0} and \eqref{con-Psi-ga} become second class. Introducing the Dirac bracket for them
allows one to eliminate all fermionic variables except $\hat\Psi^{i\alpha}{}_{b}{}^{c}$
defined in \eqref{tr-hat-X-Psi}. The Dirac bracket is equal to
\be\label{DB-Psi-0}
\left\lbrace \hat\Psi_{i\alpha}{}_{a}{}^{b},\hat\Psi^{j\beta}{}_{c}{}^{d}\right\rbrace^*
=-\,i\,\delta_{i}^{j}\,\delta^{\beta}_{\alpha}\left(\delta_{a}^{d}\,
\delta_{c}^{b}-\frac{1}{n}\,\delta_{a}^{b}\,\delta_{c}^{d}\right).
\ee

As before, with respect to the bosonic second-class constraints \eqref{primary-Z}
we introduce Dirac brackets, which for the remaining variables completely coincide with Dirac brackets \eqref{DB-b}-\eqref{DB-PZb}.

In the considered case with $\displaystyle g\,{=}-\frac{1}{n}$,
the fermionic supercharges (see \eqref{Q} and \eqref{Q-conf})
\be
Q^{i\alpha} = i\,P_{a}{}^{b}\,\hat\Psi^{i\alpha}{}_{b}{}^{a}-
\varkappa\,Z^{(i}_{a}\bar{Z}^{j)b}\,\hat\Psi^{\alpha}_{j}{}_{b}{}^{a}\,,\qquad
S^{i\alpha}= i\,X_{a}{}^{b}\,\hat{\Psi}^{i\alpha}{}_{b}{}^{a}
\label{Q-1n}
\ee
and the Hamiltonian \eqref{H-rat0} together with the bosonic charges (see \eqref{bos-conf}),
\bea
&&D=-\,\frac{1}{2}\,X_{a}{}^{b}\,P_{b}{}^{a},\qquad K=\frac{1}{2}\left(X_{a}{}^{b}\,X_{b}{}^{a}-\frac{1}{n}\,X_{a}{}^{a}
\,X_{b}{}^{b}\right),\nn
    &&I^{ij}=\varkappa\,Z^{(i}_{a}\bar{Z}^{j)b}
\hat X_{a}{}^{b}
-\frac{1}{2}\,\hat\Psi^{(i}_{\alpha}{}_{a}{}^{b}\,\hat\Psi^{j)\alpha}{}_{b}{}^{a},\qquad
F^{\alpha\beta}=\frac{1}{2}\,\hat\Psi^{i(\alpha}{}_{a}{}^{b}\,
\hat\Psi^{\beta)}_{i}{}_{b}{}^{a}\,,
\eea
form the  superconformal algebra $osp(4|2)$.

Let us consider
the bosonic core of the system with the Hamiltonian \eqref{H-rat0}.

Following the same steps as in Sec. \ref{bos-core}, we
imply the gauge
\eqref{x-fix}, i.e.,
$x_a{}^b \approx 0$, $a\neq b$,
and introduce the new variables (see \eqref{new-Z-m-b})
\be
{\it z}_a^{i}=\sqrt{\varkappa\,(x_{a}-x_0)}\,Z_a^{i}\,,\qquad
\bar{{\it z}}^a_{i}=\sqrt{\varkappa\,(x_{a}-x_0)}\,\bar{Z}^a_{i}
\ee
instead of $Z_a^{i}$ and $\bar{Z}^a_{i}$.

As a result, the Hamiltonian \eqref{H-rat0} takes the form:
\bea
H_{\rm rat.} &=&\displaystyle \frac12\, \sum_a  p_a p_a
{}+\sum_{a\neq b}
\frac{J_a{}_{i}{}^j\,J_b{}_{j}{}^i}{2\,(x_a - x_b)^2 }  \nonumber\\ [6pt]
&&{}+
\sum_{a\neq b} \frac{(2n+1)\,J_a{}_{i}{}^i\,J_b{}_{j}{}^j -2J_a{}_{i}{}^j\,J_b{}_{j}{}^i}
{8n\,(x_a-x_0) (x_b-x_0)} {}+
\sum_{a} \frac{(n-1)J_a{}_{i}{}^i\,J_a{}_{j}{}^j}{8n\,(x_a-x_0)^2}
\,,
\label{Ham-CM-nnnn}
\eea
where $J_a{}_{i}{}^j$ were defined in \eqref{S-def-n-0}.
The resulting system is described also by the set of the first-class constraints:
the constraints \eqref{T-diag-f} and the constraint \eqref{constr-P0-m-b}.
In terms of the residual variables the last constraint takes the form:
\be\lb{p0-constr}
\sum\limits_a p_a \approx 0\,.
\ee

Due to the presence of the first-class constraint \eqref{p0-constr},
the center-of-mass coordinate $x_0$ is purely gauge.
Therefore, we can choose the gauge $x_0=\mathrm{const}$.
After this, the model with the Hamiltonian \eqref{Ham-CM-nnnn} is described
by relative coordinates $x_a-x_b$, in addition to the bosonic spinor variables ${{\it z}}{}_i^a$ and Grassmann variables of the full system.

Thus in the case under consideration we end up with a deformation of the $n$-particle Calogero model of the type $A_n$\,.
In addition to terms with $\sim\displaystyle \frac1{(x_a - x_b)^2}$, it also contains terms with $\displaystyle \frac1{(x_a-x_0) (x_b-x_0)}$ and
with $\displaystyle \frac1{(x_a-x_0)^2}$. These terms cannot be reduced to terms corresponding to some known root systems of types $B_n$, $C_n$ or $D_n$ (see, e.g. \cite{Poly0607, KLPS1812}).

The system with the Hamiltonian \eqref{Ham-CM-nnnn}, as well as the previously presented model with the Hamiltonian \eqref{Ham-CM},
provide deformations of U$(2)$-spin rational Calogero system which, to the best of our knowledge, were never considered before.
As a result, in the rational case, we gain new variants of $\mathcal{N}=4$ multiparticle models deforming the U$(2)$-spin
rational Calogero system. It is of interest to further study their properties at the classical and quantum cases,
including the issue of their possible integrability\footnote{We are thankful to Armen Nersessian for discussion of this point.}.

\subsubsection{Model with extra Wess-Zumino term}
Finally, let us add to the initial multiparticle superfield action \eqref{L-X-r} with $\displaystyle g\,{=}-1/n$, an extra OSp$(4|2)$-breaking WZ term of the superfields
${\cal Z}^+_a$ and  $\bar{\cal Z}^{+ a}$ similar to \eqref{ExtraWZ}:
\be
S_{\rm rat.}\ = \
\frac{1}{4} \int d\zeta_{\rm H}\,\Big[{\rm Tr}\big(\mathscr{X}^2\big) -\frac{1}{n}\,\big({\rm Tr}\,\mathscr{X}\big)^2\Big]
\ +\ \frac{is}{2}\int d\zeta_{\rm A}^{--}\,{\rm Tr}\,{\cal V}^{++} + c\int d\zeta_{\rm A}^{--}\,{\rm Tr}\left(\mathcal{Z}^{+}\bar{\mathcal{Z}}^{+}\right).\label{act+WZ}
\ee
Skipping details, let us present the bosonic Hamiltonian as a deformation of \eqref{Ham-CM-nnnn}:
\bea
H_{\rm rat.} &=&\displaystyle \frac12\, \sum_a  p_a p_a
{}+\sum_{a\neq b}
\frac{J_a{}_{i}{}^j\,J_b{}_{j}{}^i}{2\,(x_a - x_b)^2 }  \nonumber\\ [6pt]
&&{}+
\sum_{a\neq b} \frac{(2n+1)\,J_a{}_{i}{}^i\,J_b{}_{j}{}^j -2J_a{}_{i}{}^j\,J_b{}_{j}{}^i}
{8n\left(x_a-x_0-c/\varkappa\right)\left(x_b-x_0-c/\varkappa\right)} {}+
\sum_{a} \frac{(n-1)J_a{}_{i}{}^i\,J_a{}_{j}{}^j}{8n\left(x_a-x_0-c/\varkappa\right)^2}
\,,
\eea
One can see that it exhibits both deformation parameters $\varkappa$ and $c$.
In contrast to \eqref{Ham-CM-nnnn}, we can take the proper limit $\varkappa \to 0$ that kills two last terms. Thus, the action \eqref{act+WZ} at $\varkappa = 0$ leads to a $\mathcal{N}=4$ supersymmetric extension of U$(2)$-spin rational Calogero system, though superconformal symmetry is broken.


\section{Multiparticle model II: hyperbolic case}

In ref. \cite{FIL1902,F2007}, there were considered matrix ${\cal N}=4$ supersymmetric systems gauging of which led to ${\cal N}=4$ supersymmetric extensions
of the hyperbolic Calogero-Sutherland model. Here we construct a counterpart of these systems, proceeding from the indecomposable ${\cal N}=4$ multiplet.

Instead of the action \eqref{L-X-r} we consider the superfield action
\be\lb{L-X-h}
S_{\rm hyp.}\ = \ {}-
\frac{1}{2} \int d\zeta_{\rm H}\,{\rm Tr}\big(\log\mathscr{X}\big)
\ +\ \frac{is}{2}\int d\zeta_{\rm A}^{--}\,{\rm Tr}\,{\cal V}^{++}\,,
\ee
where all matrix superfields are same as in the previous Section.
The superfield Lagrangian in \eqref{L-X-h} yields the following component Lagrangian
\bea
{\cal L}_{\rm hyp.}&=&{\rm Tr}\,\bigg\lbrace\,\frac{1}{2}\,X^{-1}\nabla {X}\,X^{-1}\nabla X+\frac{i}{2}\,X^{-1}\Psi^{i\alpha}\,X^{-1}\nabla \Psi_{i\alpha}-\frac{1}{4}\,X^{-1}N^{\alpha\beta}\,X^{-1}N_{\alpha\beta}\nn
&&+\,\frac{1}{8}\,X^{-1}N^{\alpha\beta}\,X^{-1}\Psi^i_{\alpha}\,X^{-1}\Psi_{i\beta}-\frac{1}{144}\,X^{-1}\Psi^{i}_{\alpha}\,X^{-1}\Psi_{i\beta}\,X^{-1}\Psi^{j\alpha}\,X^{-1}\Psi^{\beta}_{j}\nn
&&+\,\frac{\varkappa}{2}\,X^{-1}\left[i\left(\nabla Z^{i}\,\bar{Z}_{i}-Z^{i}\,\nabla \bar{Z}_{i}\right)-\Xi^{\alpha}\bar{\Xi}_{\alpha}\right]+\frac{\varkappa}{2}\,X^{-1}\Psi^{i\alpha}X^{-1}\left(Z_{i}\bar{\Xi}_{\alpha} + \Xi_{\alpha}\bar{Z}_{i}\right)\nn
&&+\,\frac{\varkappa^2}{4}\,X^{-1}Z^{i}\bar{Z}^{j}\,X^{-1}Z_{(i}\bar{Z}_{j)}-
\frac{\varkappa}{8}\,X^{-1}\Psi^{i\alpha}\,X^{-1}Z_{(i}\bar{Z}_{j)}\,X^{-1}\Psi^{j}_{\alpha}-
s\,{A}\,\bigg\rbrace\,,
\eea
which, after elimination of the auxiliary fields
$N^{\alpha\beta}$, $\Xi^{\alpha}$ and $\bar{\Xi}_{\alpha}$, results in the following on-shell Lagrangian
\bea
{\cal L}^{\rm on-shell}_{\rm hyp.}&=&{\rm Tr}\,\bigg\lbrace\,\frac{1}{2}\,X^{-1}\nabla {X}\,X^{-1}\nabla X+\frac{i}{2}\,X^{-1}\Psi^{i\alpha}\,X^{-1}\nabla \Psi_{i\alpha}+\frac{i\varkappa}{2}\,X^{-1}\left(\nabla Z^{i}\,\bar{Z}_{i}-Z^{i}\,\nabla \bar{Z}_{i}\right)\nn
&&+\,\frac{5}{576}\,X^{-1}\Psi^{i}_{\alpha}\,X^{-1}\Psi_{i\beta}\,
X^{-1}\Psi^{j\alpha}\,X^{-1}\Psi^{\beta}_{j}-\frac{\varkappa}{2}\,
X^{-1}\Psi^{i\alpha}X^{-1}\Psi_{j\alpha}Z^{j}\bar{Z}_{i}\nn
&&-\,\frac{\varkappa}{8}\,X^{-1}\Psi^{i\alpha}\,X^{-1}Z_{(i}\bar{Z}_{j)}\,X^{-1}
\Psi^{j}_{\alpha}+\frac{\varkappa^2}{4}\,X^{-1}Z^{i}\bar{Z}^{j}\,X^{-1}Z_{(i}\bar{Z}_{j)}-
s\,{A}\,\bigg\rbrace\,.\lb{hyp-comp}
\eea

The system \eqref{hyp-comp} is ${\cal N}=4$ supersymmetric by construction: the initial invariant action  has been built by using ${\cal N}=4$ superfields and ${\cal N}=4$ invariant measures.
${\cal N}=4$ supercharges are calculated  in the standard way. They are a deformation of the ${\cal N}=4$ supercharges of \cite{FIL1902,F2007},
supplemented by additional terms depending
on the semi-dynamical variables $Z^{i}$ and $\bar{Z}_{i}$.
In order to precisely specify the system, we consider the bosonic limit of \eqref{hyp-comp}.

\subsection{Hamiltonian formulation of bosonic core}

In the bosonic sector the canonical Hamiltonian of the system \eqref{hyp-comp} has the form:
\be
H_{\rm hyp.} = \frac{1}{2}\,\big(X^{-1}\big)_{a}{}^{b}P_{b}{}^{c}\big(X^{-1}\big)_{c}{}^{d}P_{d}{}^{a}-
\frac{\varkappa^2}{4}\,\big(X^{-1}\big)_{a}{}^{b}Z_b^{(i}\bar{Z}^{j)c}\,\big(X^{-1}\big)_{c}{}^{d}Z_{id}\bar{Z}^a_{j}
 - {\rm Tr}\,\big(A\,G\big),\label{H-hyp}
\ee
where
\be\label{Gab-hyp}
G_{a}{}^{b} = i\left(X_{a}{}^{c}P_{c}{}^{b}-P_{a}{}^{c}X_{c}{}^{b}\right)-
\frac{\varkappa}{2}\left[\big(X^{-1}\big)_{a}{}^{c} Z^{i}_{c}\bar{Z}^{b}_{i}+Z^{i}_{a}\bar{Z}^{c}_{i}\big(X^{-1}\big)_{c}{}^{b}\right]-s\,\delta_{a}^{b}\,,
\ee
are the first-class constraints.

In the system \eqref{hyp-comp} the variables $Z^{i}$ and $\bar{Z}_{i}$ are semi-dynamical.
Therefore, there arise the second-class constraints
\be
P_Z\,{}_i^b - \frac{i}{2}\,\varkappa\,\bar{Z}^c_{i}\big(X^{-1}\big)_{c}{}^{b}\approx  0\,,\qquad
P_{\bar Z}\,{}_b^i + \frac{i}{2}\,\varkappa\,\big(X^{-1}\big)_{b}{}^{c} Z_c^{i}\approx  0\,.\label{pz-hyp}
\ee
To simplify these constraints, we introduce the variables \eqref{matrix-XP}, similarly to the rational case.
As in the rational case, we fix the gauge \eqref{x-fix}, \textit{i.e.} $x_a{}^b \approx 0$, $a\neq b$ for the first-class constraints $G_{a}{}^{b} \approx 0$, $a\neq b$.
Then, we eliminate $p_a{}^b$, $a\neq b$ by
\eqref{pab} in which
\be\label{Tab-hyp}
T_a{}^b = -\,\frac{\varkappa}{2}
\left(\frac{1}{x_{a}}+\frac{1}{x_{b}}\right) Z^{i}_{a}\bar{Z}^{b}_{i}\,, \qquad a\neq b\,.
\ee
Introducing new variables\footnote{Unlike \eqref{new-Z-m-b} of the rational case,
there is now $-\varkappa$ under the square root.
This substitution of $\varkappa\to -\varkappa$ has been  made in order that the Dirac brackets of the new variables
$\left\{{{\it z}}_a^i, \bar {{\it z}}^b_j \right\}^*= i\,\delta^{i}_j\,\delta^{b}_{a}$
exactly match their Dirac brackets in the rational case.}
\bea
    &&{\it z}_a^{i}=\frac{Z_a^{i}}{\sqrt{-\varkappa\,x_a}}\,,\qquad \bar{{\it z}}^a_{i}=\frac{\bar{Z}^a_{i}}{\sqrt{-\varkappa\,x_a}}
\,, \lb{z-n-var} \\ [5pt]
    && p_{{\it z}}\,{}_i^a = \sqrt{-\varkappa\,x_a}\,P_{Z}{}_i^a\,,\qquad
p_{\bar{{\it z}}}\,{}_a^i = \sqrt{-\varkappa\,x_a}\,P_{\bar Z}{}_a^i\,,
\eea
we rewrite the second-class constraints \eqref{pz-hyp} as
\bea
    &&p_{{\it z}}\,{}_i^a + \frac{i}{2}\,\bar{{\it z}}^a_{i}\approx  0\,,\qquad
p_{\bar{{\it z}}}\,{}_a^i - \frac{i}{2}\,{{\it z}}_a^{i}\approx  0\,,   \qquad \mbox{(no summation over $a$)},
\eea
that leads to Dirac brackets \eqref{Dir-br-n-n}.

After redefining the phase variables $x_a$ and $p_a$ as
\bea
    y_{a}=\log x_{a}\,,\qquad \rho_{a}=x_{a}\,p_{a}\,,\qquad \left\{y_a, \rho_b \right\}^*=\delta_{ab}\,,
\eea
and taking into account eqs. \eqref{Tab-hyp}, \eqref{z-n-var},
the Hamiltonian \eqref{H-hyp} without the last term takes the form
\be\lb{H-hyp-1}
H_{\rm hyp.}=H_{\rm CS}+\frac{1}{4}\,\sum_{a\neq b}J_a{}_{i}{}^{i}\,J_b{}_{j}{}^{j}+\frac{1}{8}\,\sum_{a}\left(J_a{}_{i}{}^{i}\right)^2\,,
\ee
where
\be\lb{H-CS}
H_{\rm CS}=\frac{1}{2}\,\sum_{a}\left(\rho_{a}\right)^2+\frac{1}{8}\,\sum_{a\neq b}\frac{J_a{}_{i}{}^{j}J_b{}_{j}{}^{i}}{\sinh^2{\frac{\left(y_{a}-y_{b}\right)}{2}}}\ .
\ee
In \eqref{H-hyp-1} and \eqref{H-CS} the ${\rm U}(2)$ generators $J_a{}_{i}{}^{j}$ were defined in \eqref{S-def-n-0}.

The Hamiltonian \eqref{H-CS} is the Hamiltonian of the U$(2)$-spin hyperbolic Calogero-Sutherland system.
Thus, the Hamiltonian \eqref{H-hyp-1} describes the U$(2)$-spin hyperbolic Calogero-Sutherland system deformed
by the last two terms, which are bilinear in the ${\rm U}(1)$ generators $J_a{}_{i}{}^{i}$.
But among the set of first-order constraints \eqref{Gab-hyp} there are remaining  constraints $G_{a}{}^{a} \approx 0$
for which the gauges were not fixed.
They have the form \eqref{T-diag-ff}, which can be written as $J_a{}_{i}{}^{i} \approx s$ for any $a$.
Therefore, the last two terms in \eqref{H-hyp-1} are constants on the surface of constraints and
so the bosonic limit of the hyperbolic ${\cal N}=4$ system under consideration amounts to the U$(2)$-spin  hyperbolic Calogero-Sutherland system.

Thus, the superfield system \eqref{L-X-h}, up to adding some constant to the Hamiltonian,  describes
$\mathcal{N}=4$ supersymmetric extension of the U$(2)$-spin  hyperbolic Calogero-Sutherland system,
similar to $\mathcal{N}=4$ superfield model of refs. \cite{FIL1902,F2007}. Once again, addition of some external WZ type superfield
actions for the ${\bf (4, 4,0)}$ superfield subsets into the initial superfield Lagrangian  is capable to essentially enlarge
the class of such systems.

\section{Summary}

In this paper, we demonstrated the effectiveness and novelty of using indecomposable \break $d=1$, ${\cal N}=4$ multiplets.

\begin{itemize}
\item
For the first time, the nonlinear indecomposable $d=1, {\cal N}=4$ off-shell supermultiplet
${\bf (1,4,3){\supset\hspace{-1.1em}+}(4,4,0)}$ was introduced. Its both component and superfield descriptions were given.
The one-particle dynamics described by this multiplet was studied. It was shown that,
with the appropriate choice of the Lagrangian, it yields $\mathcal{N}=4$ superconformal mechanics with spin variables. The specific feature of the relevant superfield and component
Lagrangians is that their ${\bf (1,4,3)}$ parts generate the WZ type actions for the ${\bf (4,4,0)}$ spin parts, with the deformation parameter $\varkappa$ as the relevant constant WZ strength.
\item
A matrix nonlinear indecomposable supermultiplet
${\bf (1,4,3){\supset\hspace{-1.1em}+}(4,4,0)}$ was defined.
Matrix models describing multiparticle systems with $\mathcal{N}=4$ supersymmetry were considered:
\begin{itemize}
\item
The new superfield matrix model was found. It is a deformation of the $n$-particle spin rational Calogero system
and has the non-trivial traceless limit.
\item
The superfield matrix model which is $\mathcal{N}=4$ supersymmetric extension of the $n$-particle spin
hyperbolic Calogero-Sutherland system was constructed.
\end{itemize}
\item
The deformation of the spin rational Calogero system was shown to be OSp$(4|2)$ superconformal.
The generators of $\mathcal{N}=4$ supersymmetry and $\mathcal{N}=4$ superconformal symmetry for the matrix system were explicitly presented.
\end{itemize}

An important advantage of dealing with the nonlinear indecomposable $\mathcal{N}=4$  supermultiplet
${\bf (1,4,3){\supset\hspace{-1.1em}+}(4,4,0)}$ studied here is the possibility of its straightforward generalization
to higher $\mathcal{N}>4$. In particular, we have two such examples in the $\mathcal{N}=8$ case \cite{IS2512},
one of which was earlier described in terms of $\mathcal{N}=4$ harmonic superfields in \cite{FI2402}.
 The main goal of \cite{FI2402} was to construct $\mathcal{N}=8$ supersymmetric single-particle system with semi-dynamical supermultiplets.
 Its generalization to multiparticle case and the subsequent problem of gauging remained open.
 The methods developed here may hopefully allow us to resolve these problems.

Another important task is to study the integrability of the multiparticle models constructed.
Though it is rather complicated to formulate the Lax representation for the rational case of
Section \ref{Sec4}, because the Hamiltonian \eqref{Ham-CM-nnnn-0} (or \eqref{Ham-CM-nnnn}) contains non-standard potential terms,
the presence of the $\mathcal{N}=4$ superconformal symmetry in this novel multiparticle system is surely positive property which
indicates that such deformed Hamiltonians are certainly distinguished.
Moreover, the deformation of the hyperbolic Hamiltonian \eqref{H-hyp} involves no dependence on the coordinates $y_a$ and includes
only the ${\rm U}(1)$ generators $J_a{}_{i}{}^{i}$. These generators commute with all U$(2)$ generators,
so they must commute with the rest of $n-1$ integrals of motion of the U$(2)$-spin Calogero-Sutherland system.
Hence, the new model in its bosonic sector is a deformation of Calogero-Sutherland system and so it is integrable in this sector.

One can also remind recent studies of superconformal quantum mechanics in the context of describing $AdS_2$ string geometries, $D$-branes and black holes (see, e.g.,
\cite{MRV2009,LNRS2011,MRSV2203,LNRS2025}).
It would be interesting to find out possible applications of the indecomposable nonlinear supermultiplets along these lines.

\section*{Acknowledgments}
The authors thank Armen Nersessian for valuable comments. S.F. thanks Sergey Krivonos for useful discussions.

\appendix
\section{Harmonic superspace}\label{AppA}
We deal with the $d=1, {\cal N}=4$ superalgebra in the standard form (see, e.g, \cite{IS2112}):
\be\label{N4algebra}
\big\{ Q^{i}_{\alpha}, Q_{j}^{\beta}\big\} = 2\,\delta^i_j\,\delta_\alpha^\beta\, H\,,
\qquad \big[H,Q^{i}_{\alpha}\big]=0\,,
\ee
where the ${\rm SU}(2)$ indices $i=1,2$ and $\alpha= 1,2$ refer to
the automorphism group ${\rm SO}(4)\sim{\rm SU}(2)_{\rm L} \times {\rm SU}(2)_{\rm R}$\,.

The harmonic $d=1, {\cal N}=4$ superspace in the analytic basis is defined by
\bea
    \zeta_{\rm H} :=\left\lbrace t_{\rm (A)}, \theta^{\pm}_{\alpha}, u^{\pm}_i\right\rbrace,\label{HSS}
\eea
where the harmonic variables satisfy the standard completeness relation
\bea
    u^+_iu^-_j - u^+_j u^-_i = \varepsilon_{ij}\,.
\eea
The coordinates transform as
\bea
    \delta \theta^{\pm}_{\alpha}= \epsilon^{\pm}_{\alpha}\,, \qquad
    \delta u^{\pm}_{i}=0,\qquad  \delta t_{\rm (A)}
    = 2i\,\epsilon^{-\alpha}\theta^{+}_{\alpha}\,, \label{HSStr}
\eea
where $\epsilon^{\pm}_{\alpha} := \epsilon^{i}_{\alpha}u^{\pm}_{i}$\,.
The analytic subspace $\zeta_{\rm A}:=\left\lbrace t_{\rm (A)}, \theta^{+}_{\alpha}, u^{\pm}_i\right\rbrace$ forms an invariant subset in
the full superspace \eqref{HSS}.
The relevant integration measures are defined as
\bea
    d\zeta_{\rm H} := \frac{1}{4}\,du\, dt_{\rm (A)} \,d\theta^{+}_{\alpha}d\theta^{+\alpha}d\theta^{-}_{\beta}d\theta^{-\beta},\qquad d\zeta^{--}_{\rm A} := \frac{1}{2}\,du\, dt_{\rm (A)} \,d\theta^{+}_{\alpha}d\theta^{+\alpha}.
\eea
The covariant derivatives preserving the Grassmann harmonic analyticity are
\bea
    &&D^{+\alpha} = \frac{\partial}{\partial \theta^{-}_{\alpha}}\,,\nn
    &&D^{++} = \partial^{++} - i\,\theta^{+}_{\alpha}\theta^{+\alpha}\frac{\partial}{\partial t_{\rm (A)}}
    + \theta^{+}_{\alpha}\frac{\partial}{\partial \theta^{-}_\alpha} \,,\nn
    &&D^0 = \partial^0 + \theta^{+}_{\alpha}\frac{\partial}{\partial \theta^{+}_{\alpha}}
- \theta^{-}_{\alpha}\frac{\partial}{\partial \theta^{-}_{\alpha}}\,.\label{D++D+}
\eea
More details on $d=1, {\cal N}=4$ harmonic superspace can be found in \cite{IL0307} (see also Appendix\,A of \cite{IS2112}).

\end{document}